\documentclass[useAMS,usenatbib]{mn2e}
\usepackage{graphicx}
\usepackage{amssymb}
\usepackage{hyperref}

\newcommand{\ion}[2]{#1$\,${\sc {#2}}}   

\title[The Br$\gamma$ in HD\,104237]{Pre-main-sequence binaries
with tidally disrupted discs: the Br$\gamma$ in HD\,104237\thanks{Based on observations made with ESO Telescopes at Paranal Observatory under programmes 081.C-0794(A,C), 083.C-0146(A), 083.C-0236(A,C), 084.C-0983(A,B,C), 084.C-0905(A).}}
\author[P.J.V.~Garcia et al.]
{P.J.V.~Garcia,$^{1}$\thanks{E-mail: pgarcia@fe.up.pt}
M.~Benisty,$^{2,3}$
C.~Dougados,$^{3}$
F.~Bacciotti,$^{4}$
J.-M.~Clausse,$^{5}$
\newauthor
F.~Massi,$^{4}$
A.~M\'erand,$^{6}$
R.~Petrov$^{5}$
and G.~Weigelt$^{7}$\\
$^{1}$Universidade do Porto, Faculdade de Engenharia, SIM Unidade FCT n. 4006, Rua Dr. Roberto Frias, s/n, P-4200-465 Porto, Portugal\\
$^{2}$Max Planck Institute for Astronomy, D-69117 Heidelberg, Germany\\
$^{3}$Institut de Plan\'etologie et Astrophysique de Grenoble, CNRS-UJF UMR 5571, 414 rue de la Piscine, 38400 St Martin d'H\'eres, France\\
$^{4}$INAF-Osservatorio Astrofisico di Arcetri, Largo E.~Fermi 5, 50125 Firenze, Italy\\
$^{5}$Laboratoire Lagrange, UMR 7293, University of Nice Sophia-Antipolis, CNRS, Observatoire de la C\^ote d'Azur, BP 4229, 06304 Nice, France\\
$^{6}$European  Southern Observatory, Casilla  19001, Santiago 19, Chile\\
$^{7}$Max-Planck-Institut f\"ur Radioastronomie, Auf dem H\"ugel 69, 53121, Bonn, Germany}

\begin{document}

\date{ }

\pagerange{\pageref{firstpage}--\pageref{lastpage}} \pubyear{2012}

\maketitle

\label{firstpage}

\begin{abstract}

Active pre-main-sequence binaries with separations of around ten stellar radii present a wealth of phenomena unobserved in common systems. The study of these objects is extended from Classical T\,Tauri stars to the Herbig Ae star HD\,104237. The primary has a mass $2.2\pm0.2~\mathrm{M}_\odot$ and  secondary $1.4\pm0.3~\mathrm{M}_\odot$.  Spectro-interferometry with the VLTI/AMBER in the \textit{K}-band continuum and the Br$\gamma$ line is presented.

It is found that the \textit{K}-band continuum squared visibilities are compatible with a circumbinary disc with a radius of $\sim0.5$~AU. However, a significant fraction ($\sim50$~per cent) of the flux is unresolved and not fully accounted by the stellar photospheres.  The stars probably don't hold circumstellar discs, in addition to the circumbinary disk, due to the combined effects of inner magnetospheric truncation and outer tidal truncation. This unresolved flux likely arises in compact structures inside the tidally disrupted circumbinary disc.

Most ($\gtrsim 90$~per cent) of the Br$\gamma$ line emission is unresolved. The line-to-continuum spectro-astrometry shifts in time, along the direction of the Ly$\alpha$ jet known to be driven by the system. The shift is anti-correlated with the Br$\gamma$ equivalent width. It is shown that the unresolved Br$\gamma$ emission cannot originate in the jet but instead is compatible with stellar emission from the orbiting binary components. The increase in the absolute value of the equivalent width of the line takes place at periastron passage; it could arise in an accretion burst, a flare or in the increase in effective size of the emission region by the interaction of the magnetospheres. The binary longitude of the ascending node is found to be $\Omega=(235\pm3)\degr$ and the orbit retrograde.

The origin of the jet is revisited. The tidal disruption of the circumstellar disks creates  difficulties to ejection models that rely on stellar magnetosphere and disc coupling. A scenario of a stellar wind collimated by a circumbinary disc wind is suggested.

\end{abstract}


\section{Introduction}

Pre-main-sequence actively accreting binaries with separations of tens of stellar radii present a wealth of phenomena  such as tidally disrupted discs, colliding magnetospheres or periodic accretion. In parallel, these objects present overall properties (such as line profiles) that are very similar to their single counterparts. This is puzzling given that the close environment is severely disrupted by the orbital movement of the system. The evidence for tidally disrupted circumbinary actively accreting discs is growing \citep[e.g.][]{Jensen1997,Andrews2011, Harris2012}, but for only very few objects there is enough information allowing the study of the effects on the several components of the pre-main-sequence star paradigm such as the magnetosphere, the disc, the winds and jets. In this article, this sample is extended by presenting a study of HD\,104237.

The HD\,104237 spectroscopic binary has parameters $P=20$~d, $a=0.22$~AU, $m_1=2.2$~M$_\odot$  \citep[][Appendix~\ref{sec:stellar-parameters}]{Bohm2004},  a luminosity $L\sim35$~L$_\odot$ \citep{vandenAncker1998} and is located at a distance of $d=(116\pm7)$~pc \citep{Perryman1997}. It presents strong variable emission in H$\alpha$ \citep{Baines2006, Bohm2006}. Using the Br$\gamma$ line,  \citet{Garcia-Lopez2006} derived an accretion rate of \mbox{$\dot\mathrm{M}_\mathrm{acc}\sim4\times10^{-8}~\mathrm{M}_\odot~\mathrm{yr}^{-1}$}. The spectral energy distribution (SED) was fitted with a circumbinary ring with an inner rim at 0.45~AU by \citet{Tatulli2007b}. This source also drives a large-scale jet detected in Ly$\alpha$ at position angle $\theta_\mathrm{jet}=332\degr$ \citep{Grady2004}.

Most of the Br$\gamma$ emission from young stars is thought to arise in the stellar magnetosphere \citep[e.g.][]{Kurosawa2011}. The report by \citet{Tatulli2007b} of HD\,104237 Br$\gamma$ originating at scales as large as the inner rim was surprising. In contrast, \citet{Kraus2008} found that the Br$\gamma$ for HD\,104237 originates in a region smaller than the continuum. In the survey of Br$\gamma$ in young stars of \citet{Eisner2010} most objects have line emission more compact than the continuum emitted by the disc rim. However, some objects do present a contribution from extended Br$\gamma$, for example, more massive Herbig Be stars \citep{Malbet2007, Weigelt2011}. The previous HD\,104237 observations had a poor $(u,v)$ coverage; the unavailability of fringe tracking resulted in low signal-to-noise ratio (SNR). This partial information could only approximately constrain the origin of the Br$\gamma$ emission. Is it originating in the magnetosphere, the gaseous disc, the inner rim or jet? Is HD\,104237 Br$\gamma$ emission somehow related to the increased activity in tight systems as found in UZ\,Tau\,E and DQ\,Tau \citep{Basri1997, Jensen2007, Kospal2011}? These are some of the questions addressed in the present article.

In Section~\ref{sec:observations}, the observations and data reduction are presented. Then we proceed with the continuum emission results in Section~\ref{sec:LR-models}, followed by the Br$\gamma$ results in Section~\ref{sec:models}. The discussion starts with the Br$\gamma$ origin in Section~\ref{sec:discussion} and then concludes on the overall system in Section~\ref{sec:accretion-ejection}.  Details on the stellar parameters used in this work and the data analysis method used to extract line-to-continuum visibilities are presented in the Appendix. Throughout the paper the notation of \citet{Glindemann2011} is used for the interferometric quantities.

\section{Observations and data reduction}
\label{sec:observations}

\begin{table*}
\caption{Log of the observations.}
\label{tab:obs}
\begin{tabular}{ccccrccccr}
\hline
Spectral & Date & Telescope & $B$ (m) & $PA$  ($\degr$) & Spectral & Date & Telescope & $B$ (m) & $PA$  ($\degr$) \\
Mode     &      & Stations  &         &                  & Mode     &      & Stations  &         &                  \\
\hline
HR & 2010 Mar. 04 (HR1) & U2-U3 & \phantom{1}35.4  &$70.0$   & LR & 2008 June 03 (LR2) & H0-G0 & 29.6  & $-69.8$ \\
   &                    & U3-U4 & \phantom{1}60.5  &$115.6$  &	  &  		           & G0-E0 & 14.9  & $-69.8$\\
MR & 2010 Feb. 07 (MR1) & G1-D0 & \phantom{1}57.1  &$-72.4$  & 	  &		               & H0-E0 & 44.4  & $-69.8$ \\
   & 2010 Feb. 10 (MR2) & K0-G1 & \phantom{1}64.7  &$-155.4$ & LR & 2009 Apr. 04 (LR3) & G1-D0 & 61.2  & $-58.6$\\
   &                    & G1-A0 & \phantom{1}82.4  &$-91.6$  &    &                    & D0-H0 & 62.6  & $68.5$\\
   &                    & K0-A0 & 125.3            & $-119.2$&    & 		           & G1-H0 & 55.1  & $6.2$\\
   & 2010 Feb. 12 (MR3) & H0-G0 & \phantom{1}30.6  &$-93.5$  & LR & 2009 May 21 (LR4)  & G1-D0 & 62.6  & $-27.6$\\
   &                    & G0-E0 & \phantom{1}15.3  &$-93.5$  &    & 		           & D0-H0 & 58.3  & $54.0$\\
LR & 2008 May 26 (LR1)  & G1-D0 & \phantom{1}59.9  &$-37.1$  &    & 		           & G1-H0 & 40.0  & $24.5$\\
   &                    & D0-H0 & \phantom{1}56.2  &$90.0 $  & LR & 2009 May 22 (LR5)  & H0-D0 & 57.1  & $-56.4$\\
   &                    & G1-H0 & \phantom{1}49.3  &$24.2$   &    & 		           & D0-A0 & 28.5  & $-56.4$\\
   &                    &       &                  &         &    & 		           & H0-A0 & 85.6  & $-56.4$\\
\hline
\end{tabular}

\medskip
\raggedright

$B$ and $PA$ are the average baseline length and position angle. In the text, the nights are referred to as night HR1 (2010  Mar. 04), night MR1
(2010 Feb. 07), night MR2 (2010 Feb. 10), night MR3 (2010 Feb. 12), night LR1 (2008 May 26), night LR2 (2008 June 03), night LR3 (2009 Apr. 04), night LR4 (2009 May 21) and night LR5 (2009 May 22).

\end{table*}

The observations were conducted at the VLTI, using the near-infrared instrument AMBER \citep{Petrov2007}, in the spectral medium-resolution mode (MR; $R\sim1500$)  and the spectral high-resolution mode (HR; $R\sim10000$), covering the Br$\gamma$ line.
The MR observations were performed with the auxiliary telescopes (ATs), the HR observations with the unit telescopes (UTs). The field of view is about 250~mas for the ATs and 57~mas for the UTs. The FINITO fringe tracker \citep{LeBouquin2008} was used throughout the observations, although it was operated at the limit of its sensitivity with the ATs. To calibrate the continuum, we also used low-resolution data (LR; $R\sim35$) obtained without fringe tracking and with the ATs.  The baseline lengths varied between 14.5 to 125.3~m, corresponding to a maximum angular resolution $\lambda/2B$ of 1.8~mas. Each measurement was interleaved by observations of a calibrator star \citep[HD\,118934, diameter $0.877\pm0.005$~mas,][]{Merand2005}. Each MR and HR data-set typically consists of 5 to 10 individual files, each with roughly 50 individual exposures (or frames). The LR data-sets consist of a sequence of 5 files of 1000 frames each.

\begin{figure}
\begin{center}
\includegraphics[width=0.8\columnwidth]{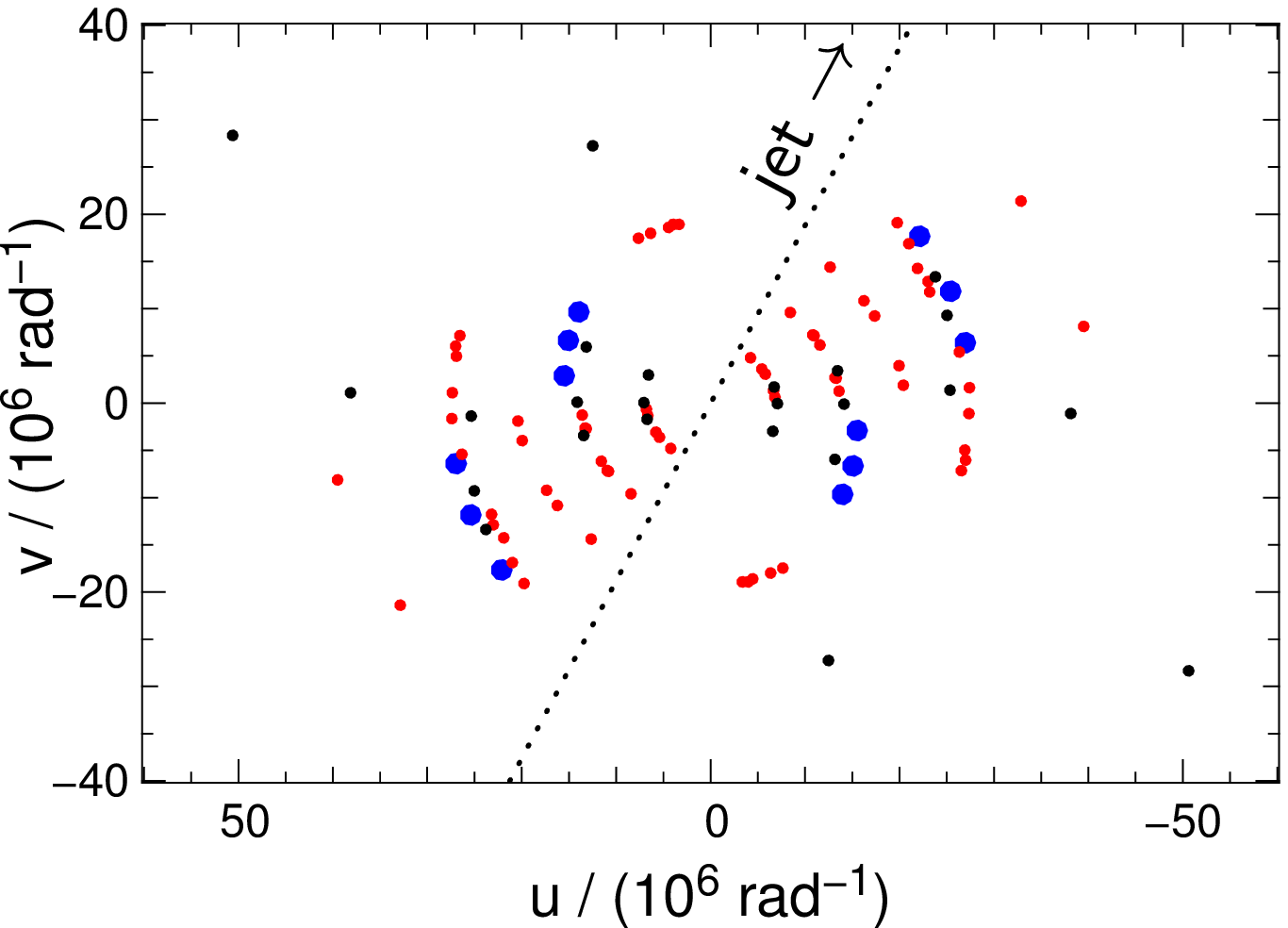}
\caption{\label{fig:UV} UV coverage of the observations, LR data (red, $\scriptscriptstyle\bullet$), MR data (black, $\scriptscriptstyle\bullet$) and HR data (blue, $\bullet$). The dashed line position angle depicts the large-scale jet direction (\mbox{$\theta_\mathrm{jet}=332\degr$}).}
\end{center}
\end{figure}

The data reduction  was  performed   using  the \textsc{amdlib} package (v3.0b), following  standard procedures described  in  \citet{Tatulli2007a}  and  \citet{Chelli2009}. Raw spectral visibilities, differential phases, and closure phases were extracted for all frames of each  observing file.  For each data set and baseline, the distributions of the fringes SNR versus optical path delay, the visibility versus SNR and histograms of the SNR were inspected. For the MR and HR data, frames with the same instrumental setup and close in time were merged to enhance the SNR. For all data-sets, a selection of 20~per cent  of the highest quality (SNR) frames  was  made. Other selections  were done without appreciable differences in final results.   The transfer   function   was  obtained   by   averaging  the   calibrator measurements, after correcting for its intrinsic diameter.  In practice, we found that the differential phase root-mean-square (rms), rather than the visibility rms, is a good indicator of the data quality. For a well-resolved object, the visibility rms will be small but not indicative of the data quality. We therefore applied a second criterion for data selection: only data with a rms in the differential phase smaller than $2\pi/50$~rad was kept. After data reduction, we used a custom software to calibrate the wavelength grid \citep{Merand2010}. The accuracy of the wavelength/velocity is $\lesssim 50$\,km~s$^{-1}$. A summary of the retained data is presented in Table~\ref{tab:obs}. The UV coverage of the final data set is presented in Fig.\ref{fig:UV}.

\begin{figure*}
\begin{center}
\includegraphics[angle=-90, width=1.6\columnwidth]{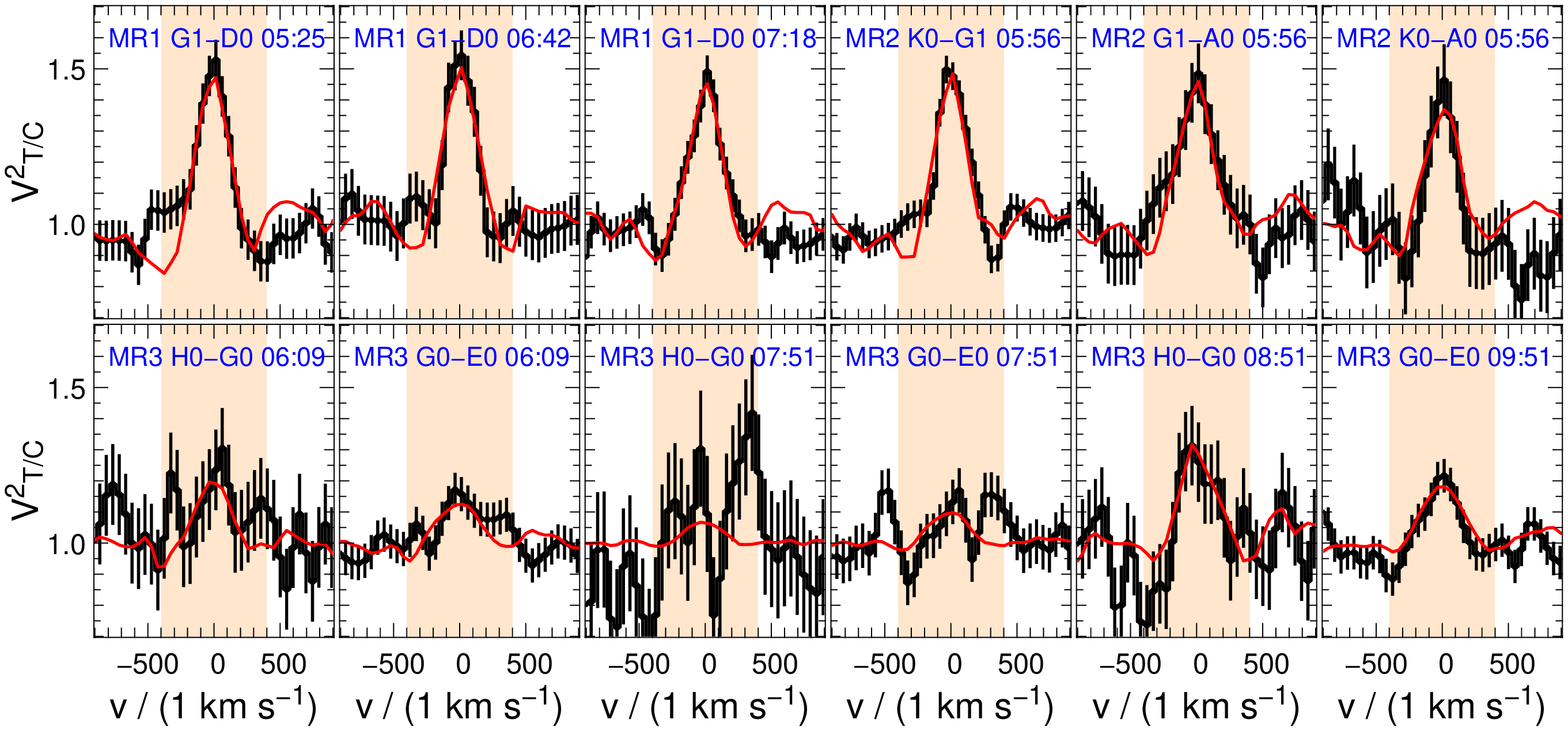}\\[5pt]
\includegraphics[angle=-90, width=1.6\columnwidth]{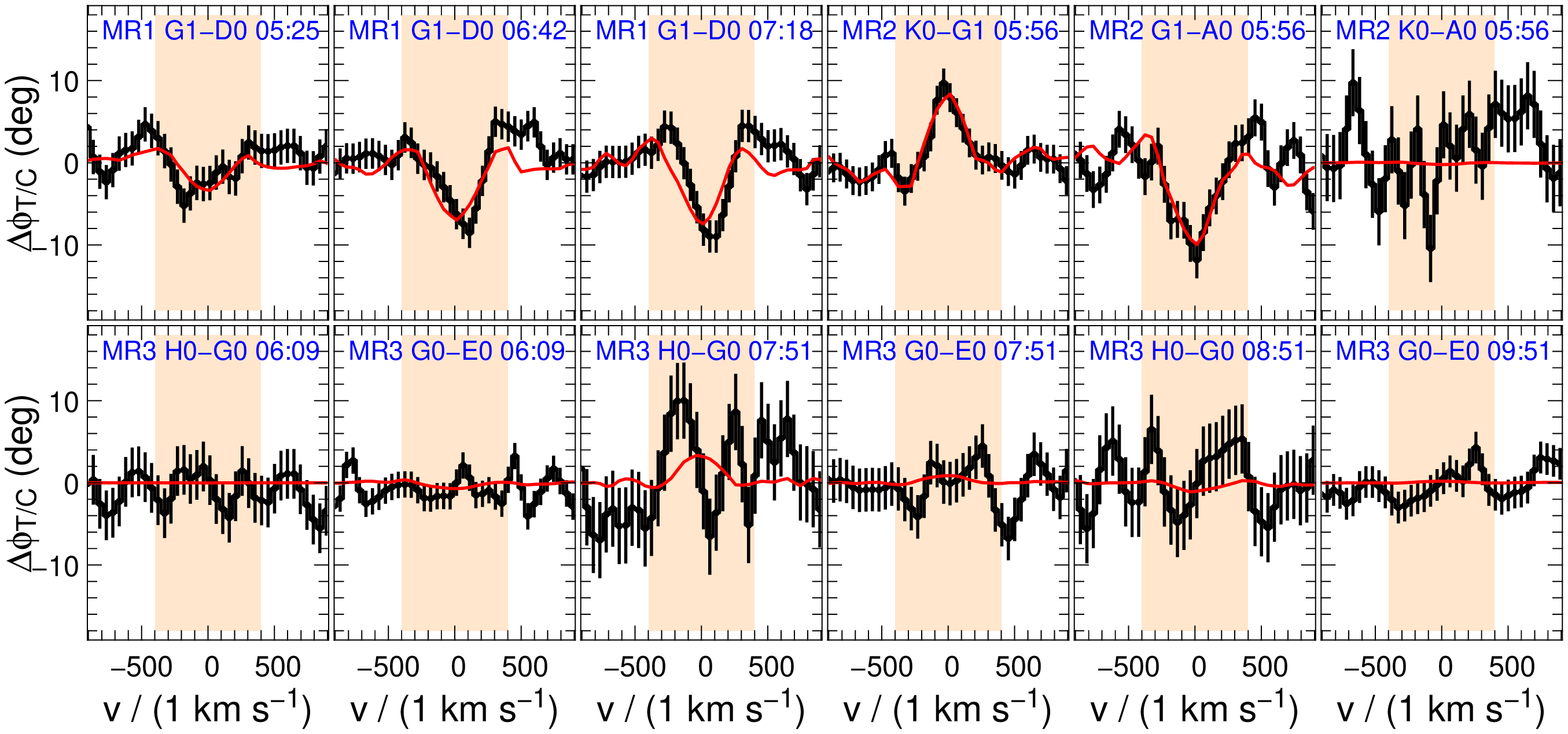}

\caption{\label{fig:data+fit+MR} MR velocity-corrected data (black) and best model fit (red, cf. Section~\ref{sec:brgamma} and Appendix~\ref{sec:least-squares}). The ros\'e background is the region used in the fit, the blue legend contains the night code, baseline and hh:mm of the data. Top: continuum-normalized squared visibility ($V^2_\mathrm{T/C}$); bottom: differential phase ($\Delta\phi_{\rm T/C}$).  Appendix~\ref{sec:fit} discusses the fit quality and in particular  the data set MR3~H0-G0~07:51.}
\end{center}
\end{figure*}

\begin{figure*}
\begin{center}
\includegraphics[angle=-90, width=1.6\columnwidth]{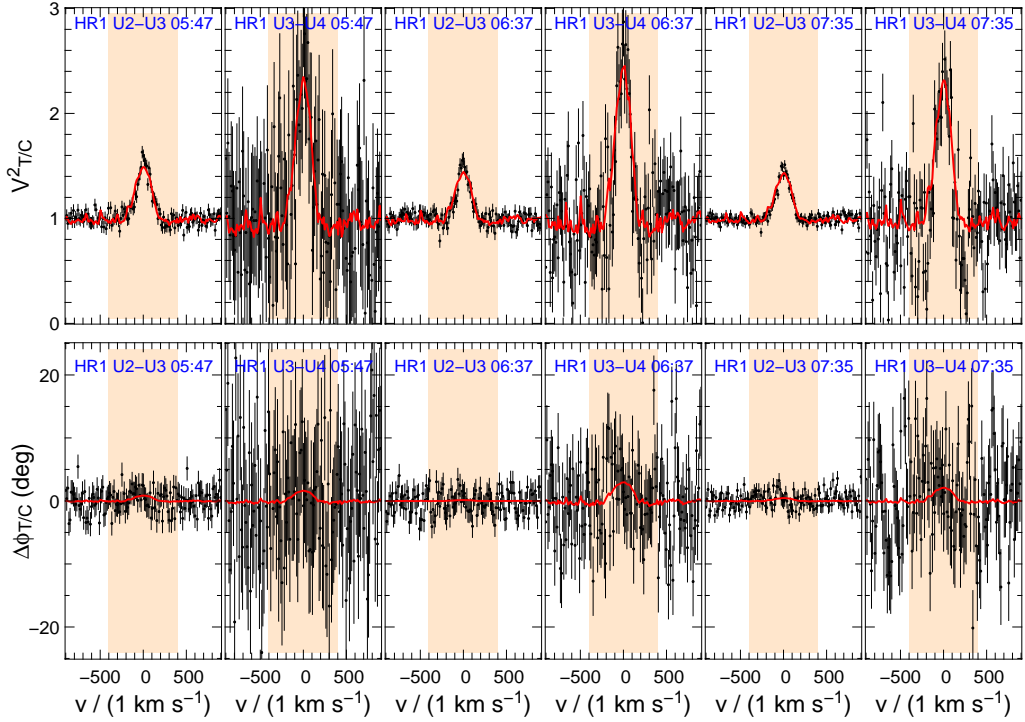}

\caption{\label{fig:data+fit+HR} HR velocity-corrected data (black) and best model fit (red, cf. Section~\ref{sec:brgamma} and Appendix~\ref{sec:least-squares}). The ros\'e background is the region used in the fit, the blue legend contains the night code, baseline and hh:mm of the data. Top: continuum-normalized squared visibility ($V^2_\mathrm{T/C}$); bottom: differential phase ($\Delta\phi_{\rm T/C}$).}
\end{center}
\end{figure*}

The absolute value of the visibilities obtained with the UT baselines could not be determined due to random vibrations of the telescopes and/or different performance of the fringe tracker during the target and calibrator observations. We also found some discrepancies between the absolute values in the continuum visibilities retrieved from the MR/AT observations and from the LR/AT data. This issue affects all spectral channels in the same way and does not change any conclusion relative to the spectral analysis. The result of the MR/HR data reduction is therefore squared visibilities normalized to the continuum ($V^2_{\rm T/C}=V^2_\mathrm{T}/V^2_\mathrm{C}$) and differential phases relative to the continuum ($\Delta \phi_{\rm T/C}=\phi_\mathrm{T}-\phi_\mathrm{C}$). The final data-set is presented in Fig.~\ref{fig:data+fit+MR} for the MR/AT data and Fig.~\ref{fig:data+fit+HR} for the HR/UT data.

\begin{figure}
\begin{center}
\includegraphics[width=5cm]{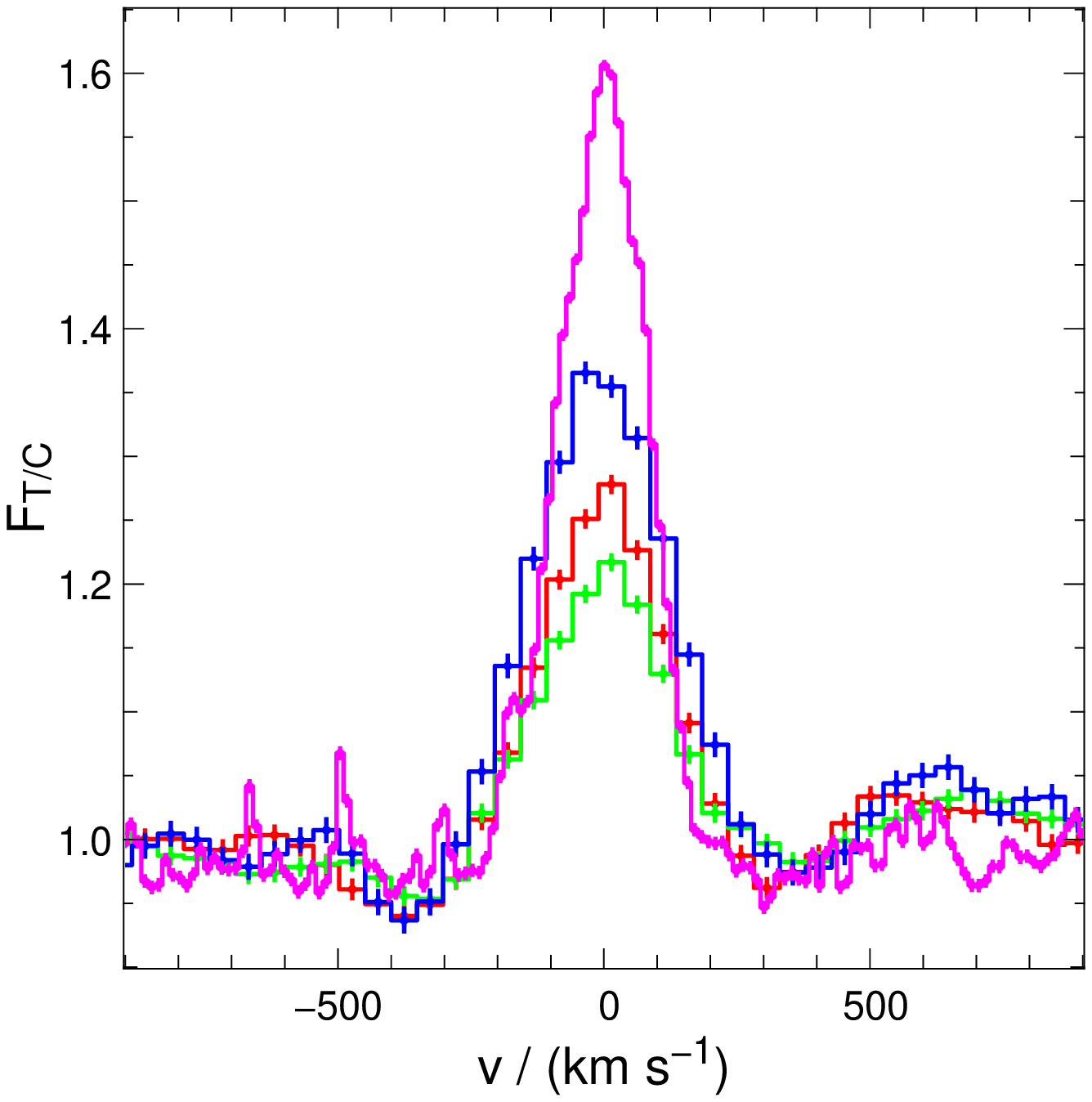}
\caption{\label{fig:spectra} Spectroscopic observations for each night. Medium-resolution: night MR1 (red), night MR2 (green), night MR3 (blue) and high-resolution: night HR1 (magenta).}
\end{center}
\end{figure}

In Fig.~\ref{fig:spectra}, the Br$\gamma$ spectra are presented. The data for each night was continuum normalized and averaged.  The first three nights spectra (MR1, MR2 and MR3) were obtained with medium spectral resolution, the fourth night spectrum (HR1) with high spectral resolution.  The line intensity increases with time but the profile shape remains very similar.

\section{The continuum emission of HD\,104237}
\label{sec:LR-models}

\begin{figure}
\begin{center}
\includegraphics[angle=0, width=5.5cm]{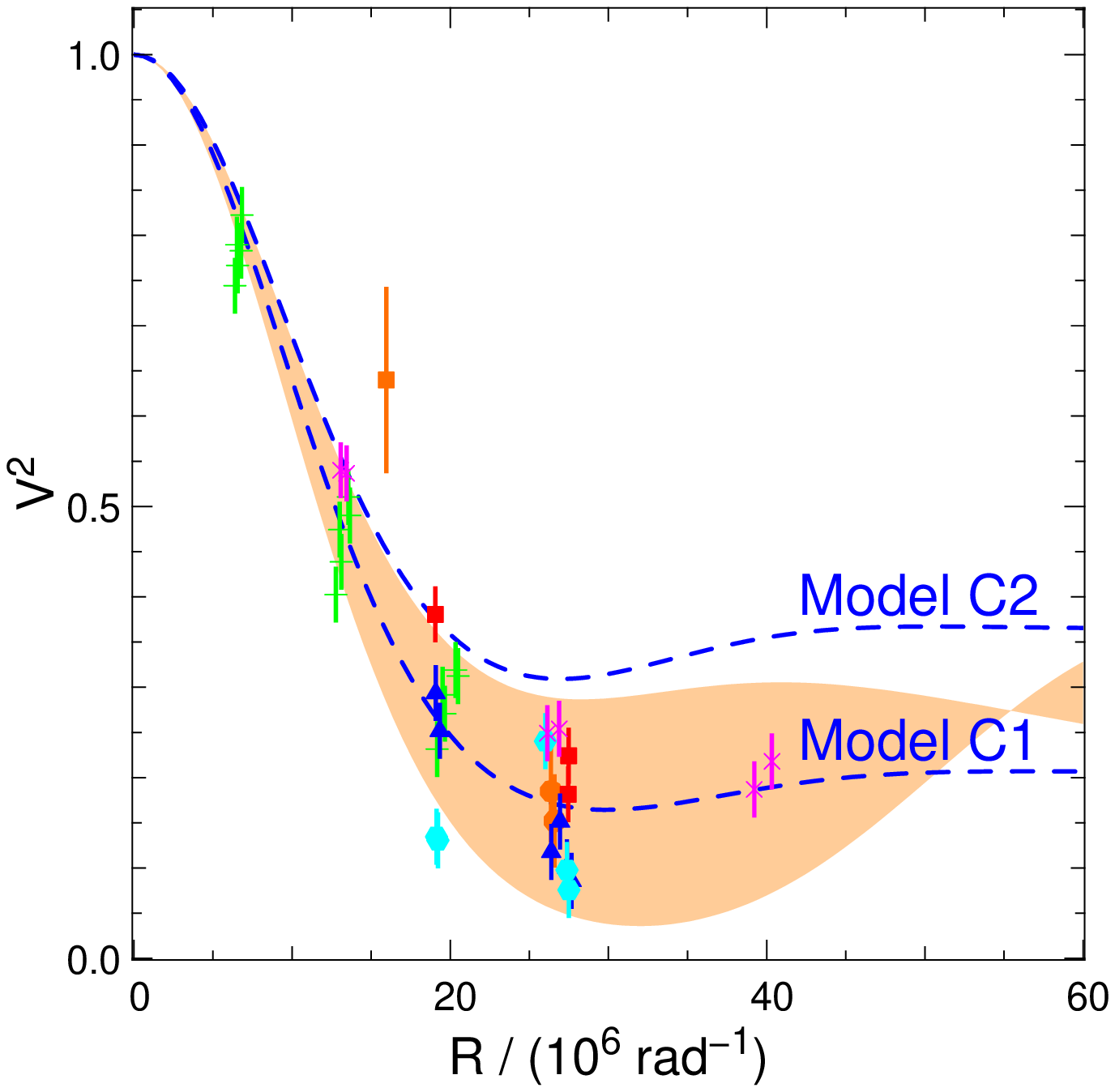}\\[10pt]
\includegraphics[angle=0, width=5.5cm]{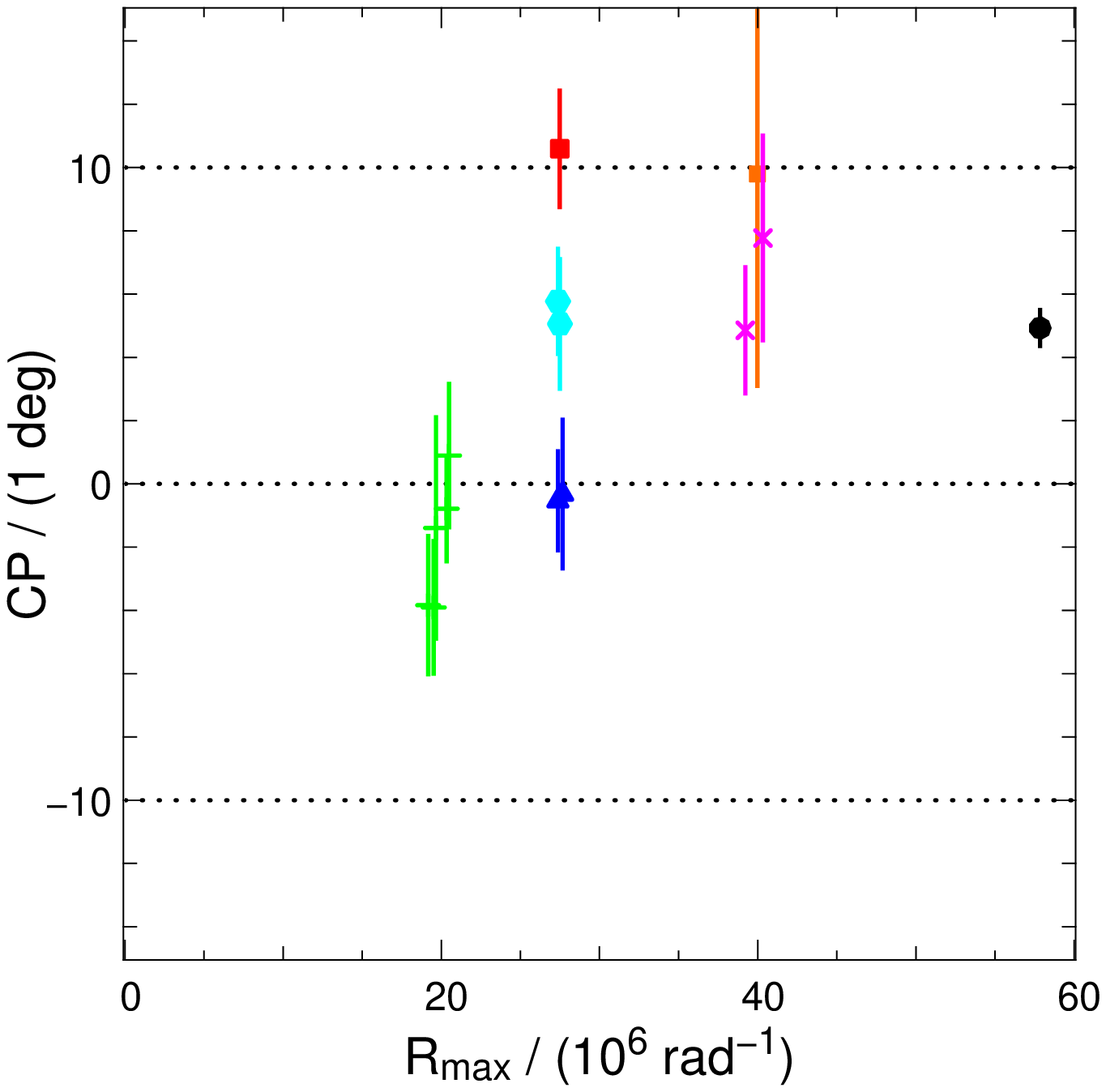}
\caption{\label{fig:continuum} \textit{K}-band continuum data. Top: $V^2$ versus spatial frequency radius $R$ in the ($u,v$) plane. The  Model~C1 line is the best fit to the data. The ros\'e range and Model~C2 depict the effect of the binary (cf. Section~\ref{sec:LR-models} for details).  Bottom: closure phase (CP) versus maximum triangle spatial frequency radius ($R_\mathrm{max}$). The data colour codes are: night LR1 (red, $\scriptscriptstyle\blacksquare$), night LR2 (green, $+$), night LR3 (blue, $\blacktriangle$), night LR4 (cyan, $\bullet$), night LR5 (magenta, $\times$), night MR2 (black, $\bullet$ in closure phase only). \citeauthor{Kraus2008} data is plotted in orange.}
\end{center}
\end{figure}

The \textit{K}-band continuum data consists of LR squared visibilities ($V^2$) and LR and MR closure phases. The MR and HR $V^2$ continuum values are unavailable due to transfer function variability. Only one set of MR data allows the computation of the closure phase. The data is presented in Fig.~\ref{fig:continuum}. In the following the best fit geometric model to the data is presented, then the scatter in the data is explained with the binarity of the object.

\begin{table}
\centering
\caption{Continuum $V^2$ models and derived parameters. }
\label{tab:LR-model}
\begin{tabular}{lccc}
\hline
Model               & Value (mas) & $f$ (per cent) &$\chi^2_\nu$\\\hline
Model~C1 &$r=3.5\pm0.8$, $\sigma=1.8\pm0.6$ &$45\pm7$ &5.4\\
Model~C2 &$r=4.0$, $\sigma=1.8$ &$60$ &n.a.\\
\hline
\end{tabular}

\medskip
\raggedright
Models C1 and C2: ring of radius $r$ with radial Gaussian $\sigma$ cross-section and point source with fractional flux $f$; $\chi^2_\nu$ is the reduced $\chi^2$.

\end{table}

\paragraph*{Geometric model}
In the astrophysical context of HD\,104237, the $V^2$ continuum data was fitted with an extended ring concentric with a point source (Model~C1). The extended ring has a radius $r$ and radial Gaussian ring cross-section. It mimics an unresolved spectroscopic binary surrounded by an extended circumbinary sublimation rim.  The best-fitting parameters and reduced $\chi^2$ are presented in Table~\ref{tab:LR-model} and the data and fit in Fig.~\ref{fig:continuum}. The fit presents a significant $\chi^2_\nu$ and should be taken as a rough estimate of the circumbinary disc size and unresolved flux fraction. The estimated disk size from model fitting is model dependent. Using other models we found it to vary from 3.0 to 4.4~mas. A result common to all models is that the fractional flux found in the point component is larger than the 30~per cent expected from the SED fits to the disc by \citet{Tatulli2007b}.

\paragraph*{Effect of the spectroscopic binary}
Model~C1 predicts zero closure phase. The binary asymmetric emission generates a non-zero closure phase. To explain the closure phase the unresolved flux of Model~C1 was allocated to both components of the binary (described in Table~\ref{tab:radial-velocity}). Fitting the closure phase with such a model had \mbox{$3 \lesssim \chi^2_\nu \lesssim 5$}, which supports the spectroscopic binary as causing most of the observed closure phase values and scatter.\\
With regards to the $V^2$, the effect of including a binary is to decrease the visibility. The projected time averaged binary separation in the plane of the sky is $(2.2\pm0.7)~\mathrm{mas}$. The binary is marginally resolved at the longer baselines.  In Fig.~\ref{fig:continuum} (Top) the effect of including a binary in the modeling is illustrated. Model~C2 has an increased radius $r=4.0$~mas and increased unresolved flux fraction $f=60$~per cent. Apparently this model does not fit the data. However, if the unresolved flux of Model~C2 is allocated to the binary then the resulting range of $V^2$ values (ros\'e range in the Figure) covers most of the data points. The binary is a satisfactory explanation to the observed scatter in the $V^2$ data.

\paragraph*{Summary of the fits}
The continuum interferometric data of HD\,104237 is reasonably well explained by  a spectroscopic binary surrounded by a circumbinary disc of radius $\sim4$~mas. This value is intermediate to the 3.9~mas derived by \citet{Tatulli2007b} and 5.0~mas found by \citet{Kraus2008}. The flux allocated to the circumbinary disc is smaller ($f\sim50$~per cent) than the \textit{K}-band excess ($f\sim70$~per cent) derived from the SED fits by \citet{Tatulli2007b}. Significant excess exists in unresolved emission in all models. The continuum emission will be discussed in Section~\ref{sec:inner_au}.

\section{The Br$\gamma$ complex visibilities}\label{sec:brgamma}\label{sec:models}

\subsection{Line-to-continuum observables}

\begin{figure}
\begin{center}
\includegraphics[width=5.5cm]{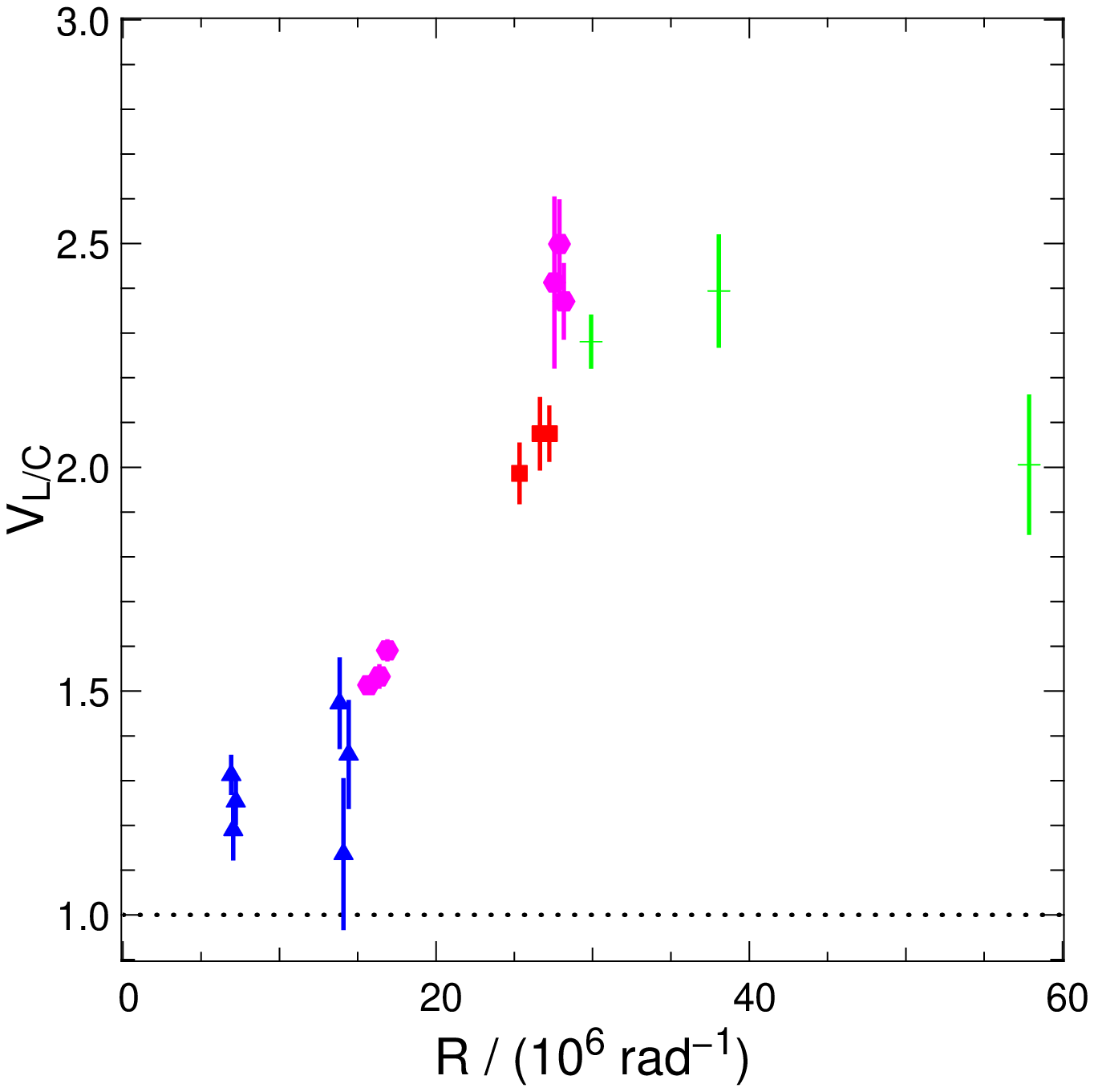}\\[10pt]
\includegraphics[width=5.5cm]{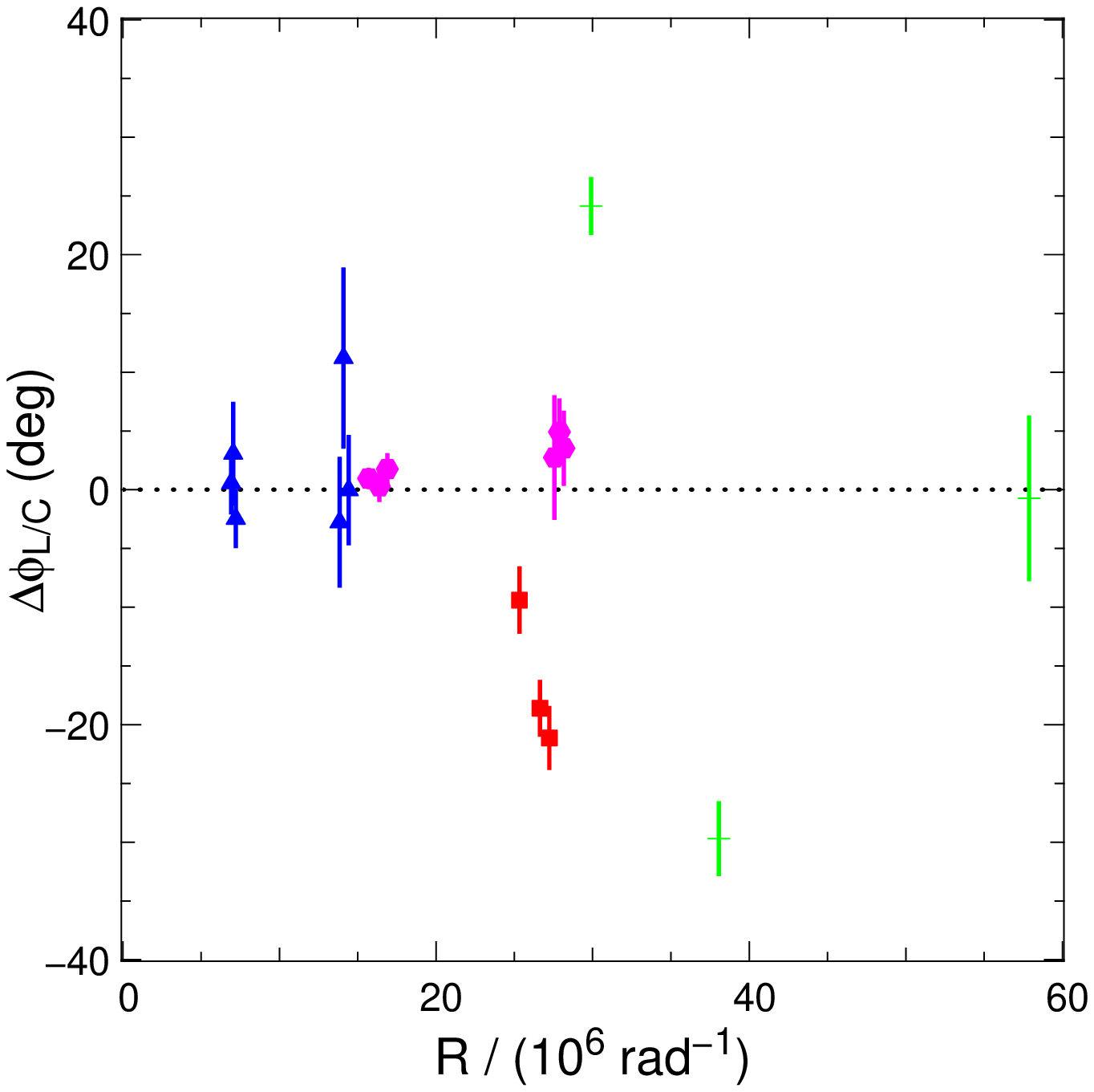}
\caption{\label{fig:fit-MR+HR} Extracted quantities from the least-squares deconvolution for each baseline versus frequency radius $R$ in the ($u,v$) plane. Line-to-continuum visibility ratio $V_{\rm L/C}$ (top) and line-to-continuum differential phase $\Delta\phi_{\rm L/C}$ (bottom). Note that $V_{\rm L/C}$ is a ratio and can be larger than 1. MR data of night MR1 (red, $\scriptscriptstyle\blacksquare$), night MR2 (green, $+$) and night MR3 (blue, $\blacktriangle$). HR data of night HR1 (magenta, $\bullet$).}
\end{center}
\end{figure}

As the SNR of the data is relatively low and the line marginally spectrally resolved in MR, a special procedure was developed to extract the line-to-continuum visibilities and phases. This procedure called least-squares deconvolution is presented in detail in Appendix~\ref{sec:least-squares}. It provides a factor of two increase in the precision of the extracted line-to-continuum squared visibility and differential phase. In Fig.~\ref{fig:fit-MR+HR} the extracted quantities are presented. The overall results are that the Br$\gamma$ is more compact than the continuum $V_\mathrm{L/C}>1$ and that the line centroid presents sometimes a displacement with respect to the continuum $\Delta\phi_\mathrm{L/C}\neq 0$. The line-to-continuum closure phase for the MR2 night, where a complete triangle was available, is found to be zero within errors $\mathrm{CP}=(0\pm7)\degr$, i.e. in the line-to-continuum spatial asymmetry ("skewness") is the same.

\subsection{Spectro-astrometry}

\begin{figure}
\begin{center}
  \includegraphics[height=5.5cm]{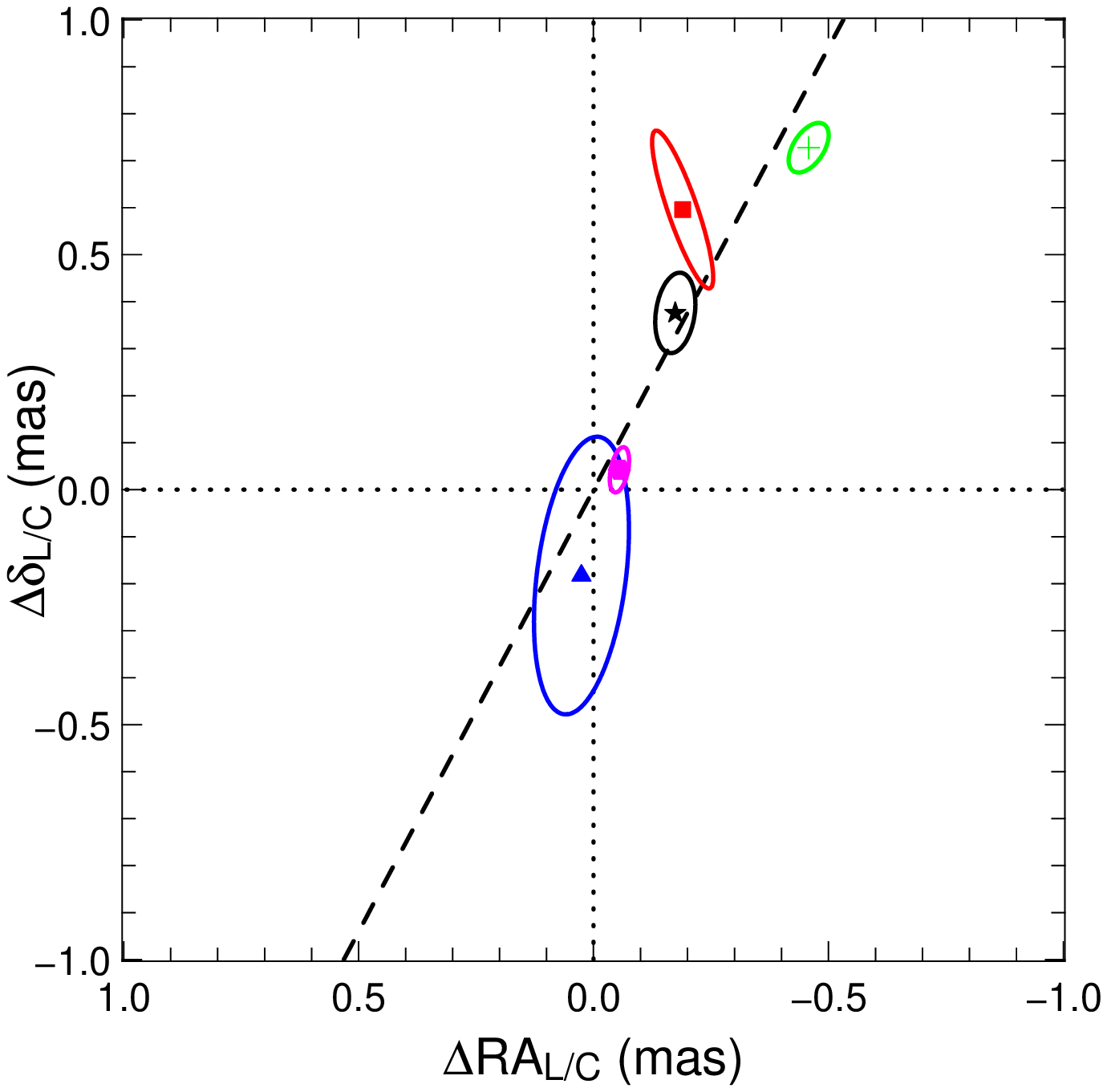}\\[10pt]
  \includegraphics[height=5.5cm]{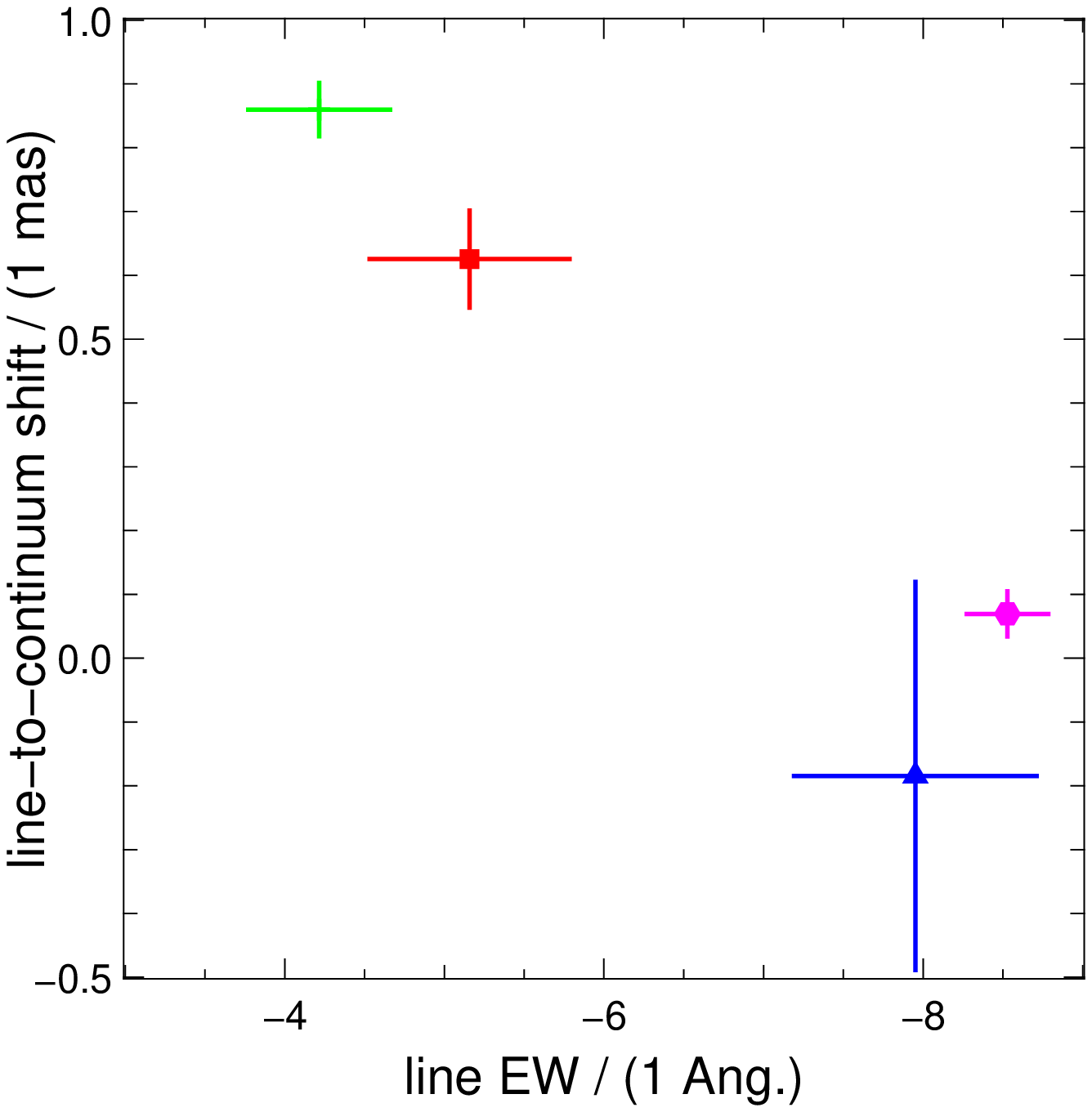}
\caption{\label{fig:astrometry} Br$\gamma$ spectro-astrometry. The continuum is assumed at the origin. Top: average position (relative declination   $\Delta\delta_\mathrm{L/C}$ and right ascension $\Delta$RA$_\mathrm{L/C}$) for all nights' data (black, $\bigstar$); for the individual nights the coding is the same as in Fig.~\ref{fig:fit-MR+HR}. The ellipses present the $1\sigma$ error in the spectro-astrometry. The dashed straight line shows the jet position angle. Bottom: shift along jet axis versus Br$\gamma$ equivalent width.}
\end{center}
\end{figure}

The shift property of the Fourier transform relates the measured wavelength-differential phase to the spectro-astrometric centre of the source brightness distribution. In particular, the line-to-continuum phase $\Delta \phi_{\rm L/C}$  is related to the line-to-continuum spectro-astrometry $\bmath{\alpha}_{\rm  L/C}$  via
\[\Delta \phi_{\rm L/C}=-2\pi \bmath{\alpha}_{\rm L/C}\cdot\bmath{R},\]
with $\bmath{R}$ the position vector in the $(u,v)$ plane.  HD\,104237 is a spectroscopic binary with a period of 20~d. Therefore, fits for each night are presented in Fig.~\ref{fig:astrometry} (top). Except for the lowest angular resolution data (smallest baselines), the measured shifts are displaced from the origin by more than $3\sigma$. Because these are line-to-continuum positions the observed shift can be caused either by a continuum shift, or by a line shift or by both. The reader is reminded that in Section~\ref{sec:LR-models} it was found that the continuum had a non-zero closure phase, therefore its astrometric centre is not at the origin.

Fig.~\ref{fig:astrometry} (top) shows that the spectro-astrometric position changes from night to night along the Ly$\alpha$ jet direction. Comparison with the integrated spectra (Fig.~\ref{fig:astrometry}, bottom), which are also variable with time (cf. Fig.~\ref{fig:spectra}) shows that the shift
magnitude correlates with the Br$\gamma$ equivalent width. Larger Br$\gamma$-to-continuum flux translates in smaller shifts.

\subsection{Complex visibility}

\begin{figure}
\begin{center}
  \includegraphics[width=5.5cm]{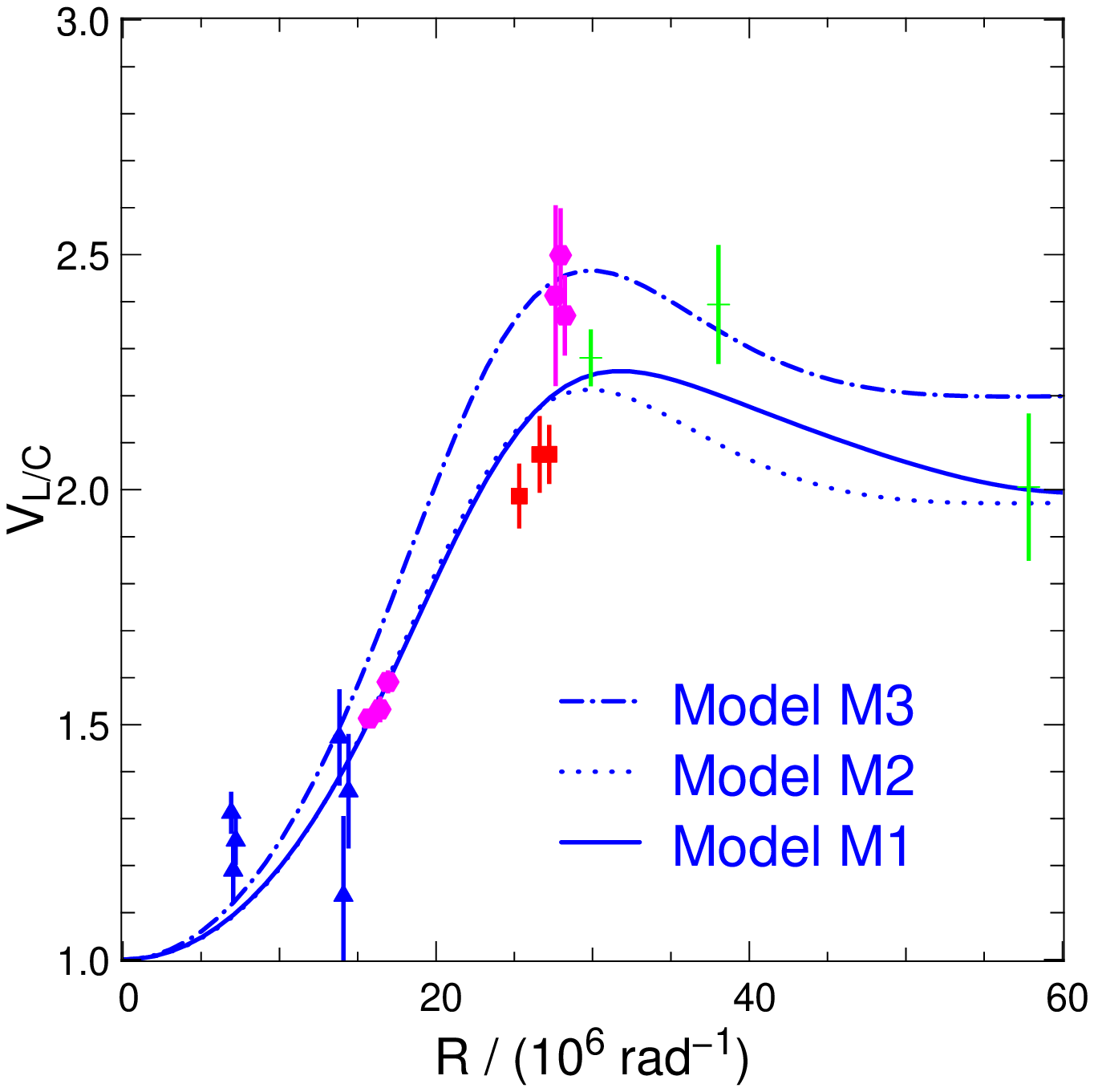}\\[10pt]
  \includegraphics[width=5.5cm]{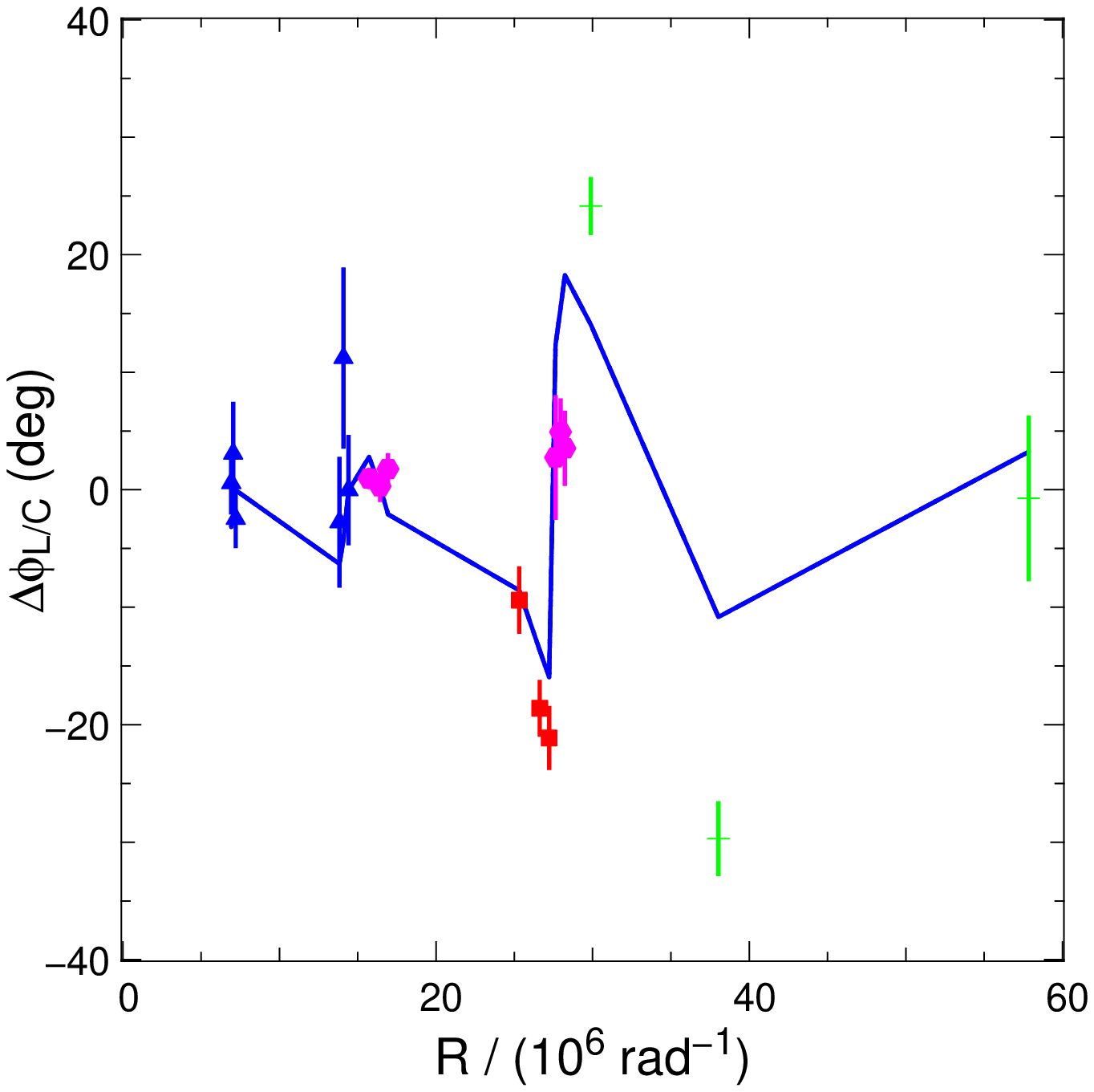}
\caption{\label{fig:complex-fit} Line-to-continuum complex visibility models. The lines are the best-fitting models to the data. The fit information is presented in Table~\ref{tab:complex-fit}. For $\Delta\phi_\mathrm{L/C}$ the lines overlap. The data coding is the same as in Fig.~\ref{fig:fit-MR+HR}. }
\end{center}
\end{figure}

\begin{table*}
\centering
\caption{Line-to-continuum complex visibility models and derived parameters.}
\label{tab:complex-fit}
\begin{tabular}{lccccc}
\hline
 model             & rel. $\delta$ (mas)& rel. $\alpha$ (mas) & (mas) & $f$ (per cent) &$\chi^2_\nu$\\\hline
Model~M1 &$-0.17\pm0.04$ &$0.38\pm0.08$ &$r= 5.4\pm1.0$ &$92\pm1$ &6.7\\
Model~M2 &$-0.17\pm0.04$ &$0.38\pm0.08$ &$\sigma=3.5\pm1.6$ &$90\pm3$&6.9\\
Model~M3 &$-0.17\pm0.06$ &$0.38\pm0.11$ &- &100 &14\\
\hline
\end{tabular}

\medskip
\raggedright

Model~M1: displaced $\delta$-ring of radius $r$ and point; Model~M2: displaced Gaussian and point source; Model~M3: point source. $f$ is the fraction of total flux in the point source. The continuum model used is Model~C1 from Table~\ref{tab:LR-model}.

\end{table*}

The line-to-continuum wavelength differential phase and visibility can be fitted simultaneously. The model for the complex visibility is
\[V_\mathrm{L/C} = \exp(-2\pi i \bmath{\alpha}_\mathrm{L/C}\cdot\bmath{R}) \frac{V_\mathrm{L}}{V_\mathrm{C}(\mathrm{Model\,C1})}.\]
It consists on a displacement $\alpha_\mathrm{L/C}$ and a centre-symmetric visibility $V_\mathrm{L}$ whose functional form is described in Table~\ref{tab:complex-fit}. Model~C1 is used to calibrate the data, as it is centre-symmetric it has a zero closure phase. This model is not necessarily optimal because, as shown in Section~\ref{sec:LR-models}, the continuum has a non-zero closure phase. However, a calibration of the line closure-phases in the same way as was done for the visibility would add too many parameters and make the result even more model dependent. For a given night only a small number of $(u,v)$ points is available and there is poor baseline distribution. Therefore, a night-by-night fit of the data was not attempted. With this caveat in mind, the fits are presented in Table~\ref{tab:complex-fit} and Fig.~\ref{fig:complex-fit}. The main result is that the Br$\gamma$ emission is composed of an unresolved component accounting for at least of $\sim 90$~per cent of the flux. Given the uncertainties in the continuum visibility calibration, a model where all the Br$\gamma$ flux is unresolved (Model~M3) is also roughly compatible with the data. The size of the extended remaining Br$\gamma$ component is compatible with the continuum fit presented in Section~\ref{sec:LR-models}, but this result is strongly dependent on the actual geometric model used. We therefore abstain from over-interpreting this ``resolved'' component of Br$\gamma$ obtained in the models of Table~\ref{tab:complex-fit}. The models also present an astrometric signature at the position of the average spectro-astrometry.

\subsection{Summary of the modeling}

The results of the modeling can be summarized as follows:
\begin{itemize}
\item[-] The Br$\gamma$-to-continuum spectro-astrometry position moves along the jet direction.
\item[-] The Br$\gamma$-to-continuum spectro-astrometry position changes with time.
\item[-] The Br$\gamma$-to-continuum spectro-astrometry movement is anti-correlated with the Br$\gamma$ equivalent width  absolute value.
\item[-] Most ($\gtrsim90$~per cent) of the Br$\gamma$ emission is unresolved.
\item[-] The remaining extended Br$\gamma$ flux is compatible in size with the continuum but this result is model dependent.
\end{itemize}


\section{The origin of the Br$\gamma$ emission}
\label{sec:discussion}

In this section, we argue that the compact Br$\gamma$ emission observed along the direction of the large-scale jet cannot originate in the jet or a disc wind  but instead arises in the close stellar environment of each binary component. Mixed models of binary plus jet were not attempted due to the high number of parameters with regard to the available data.

\subsection{Jet scenario}
\label{sec:jet}

The detection of Br$\gamma$ in HD\,104237 along the direction of the Ly$\alpha$ jet could possibly point to a common origin. Jets are known to emit hydrogen recombination lines, but the Br$\gamma$ emission is normally compact and unresolved \citep[e.g.][]{Davis2011}. \citet{Beck2010} find  the Br$\gamma$ emission to have a 2-10~per cent contribution from extended components at scales $\lesssim 200$~AU. At much smaller scales, Br$\gamma$ jets were detected in the 2008 Z\,CMa outburst by \citet{Benisty2010b}, but most objects show unresolved emission \citep{Eisner2010, Kraus2008}.

\begin{figure*}
\begin{center}
  \includegraphics[width=0.6\columnwidth]{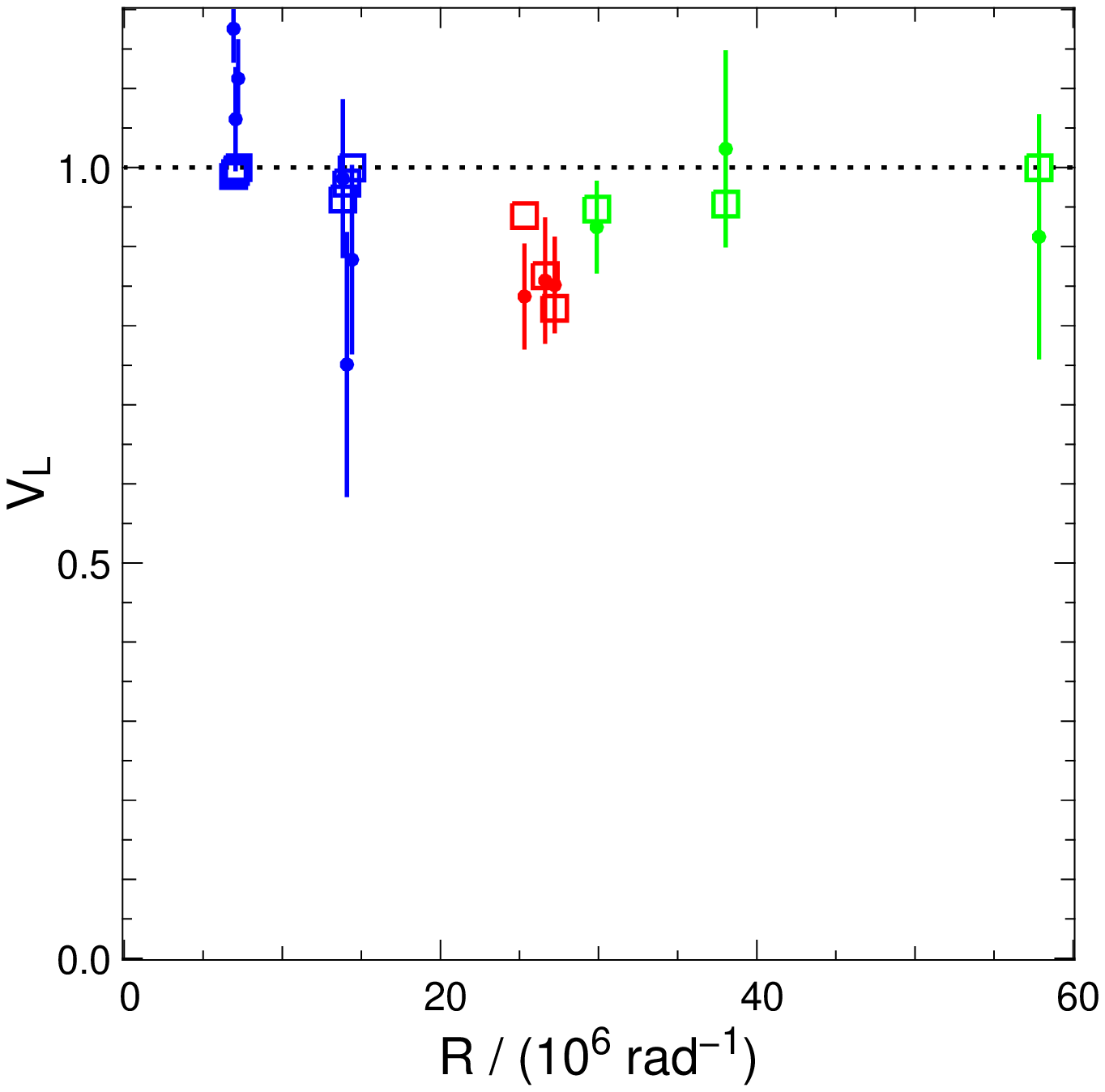}\hspace{5mm}
  \includegraphics[width=0.6\columnwidth]{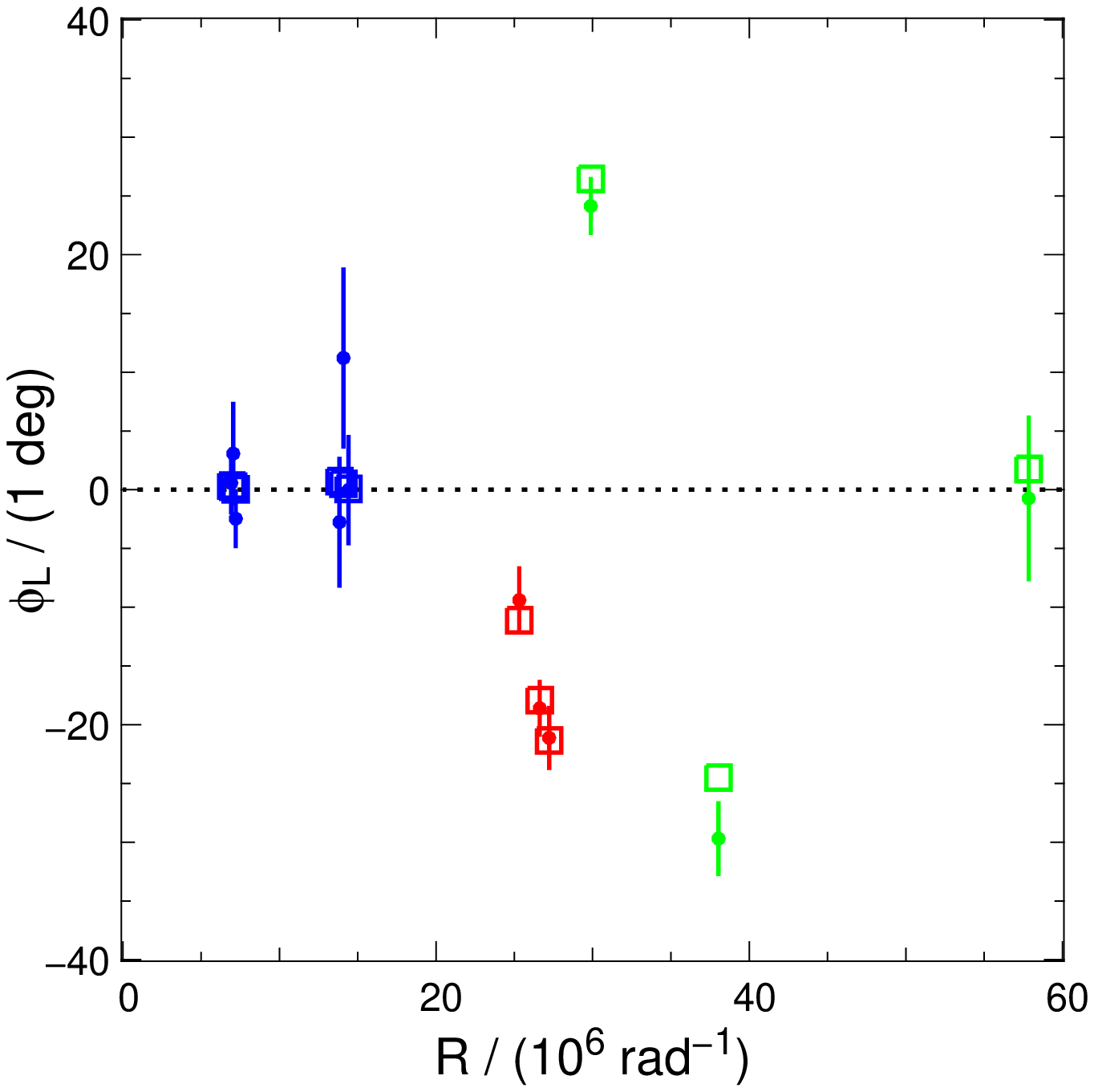}\hspace{5mm}
  \includegraphics[width=0.6\columnwidth]{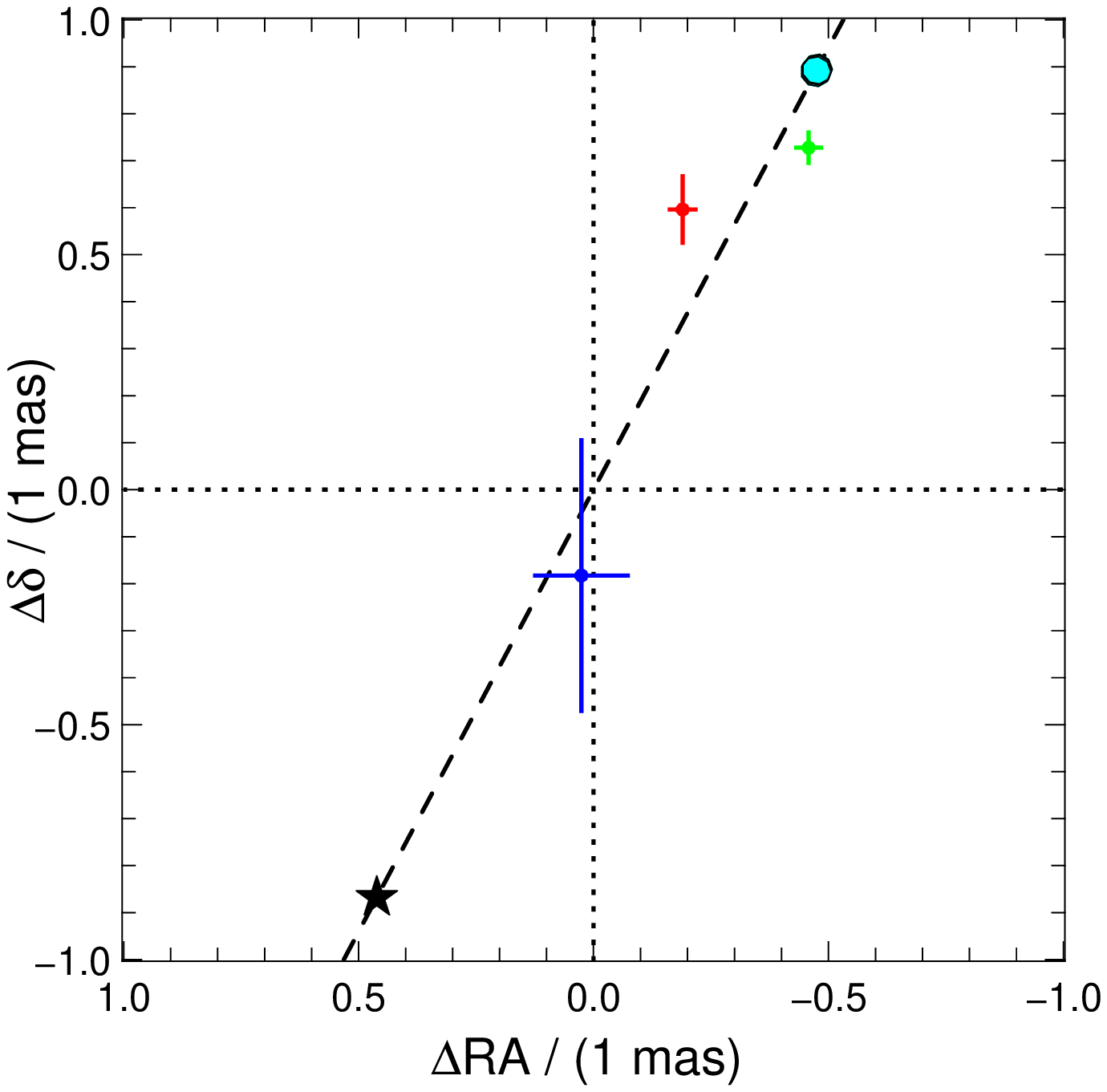}
\caption{\label{fig:astrometry-jet-fit} Jet scenario. The night
  of observation is colour encoded as in  Fig.~\ref{fig:spectra}.   Line visibility (left) and line phase (centre)
   versus   spatial frequency radius $R$ in the ($u,v$) plane. The open squares are the best-fitting model.
   Right: drawing of the best-fitting model. The star
  ($\star$) is the stellar position, the filled circle is the knot position,
  the crossed points the astrometric data.}
\end{center}
\end{figure*}

\paragraph*{Model} The Br$\gamma$ is modeled with a mixed contribution to the line from a displaced unresolved knot ($F_\mathrm{knot}$) and from variable stellar emission ($F_\star$) at a fixed position $\bmath{r}_\star$. The observed position ($\bmath{p}$) of the Br$\gamma$ is
\[\bmath{p}=\frac{F_\star}{F_\mathrm{T}} \bmath{r}_\star + \frac{F_\mathrm{knot}}{F_\mathrm{T}}\bmath{r}_\mathrm{knot},\]
with $F_\mathrm{T}= F_\star + F_\mathrm{knot}$ and $\bmath{r}_\mathrm{knot}=d_\mathrm{knot}\bmath{u}_\mathrm{jet}$ directed along the jet position angle ($\theta_\mathrm{jet}=332^\circ$). In Fig.~\ref{fig:astrometry} (top) the points MR1 and MR2 are located along the jet position angle, the point MR3 is compatible with a position at the origin. To explain this last point $\bmath{r}_\star$ must be located opposite to the jet direction. The position angle of the star is therefore fixed at $\theta_\mathrm{jet}-\pi$, leaving free only $d_\star$ ($\bmath{r}_\star=-d_\star \bmath{u}_\mathrm{jet}$). The knot moves with speed $v_\mathrm{knot}$, its distance is \mbox{$d_\mathrm{knot}=d_{\mathrm{knot}, t_0} + v_\mathrm{knot}(t-t_0)$}.

\paragraph*{Visibility calibration} The Br$\gamma$ visibilities and phases are recovered making use of the continuum Model~C1. The line visibility ($V_\mathrm{L}$) then becomes
\begin{equation}\label{eq:VL_calib}
V_\mathrm{L}=V_\mathrm{L/C} \, V_\mathrm{C}(\mathrm{Model \,C1})
\end{equation}
and the line phase ($\phi_\mathrm{L}$)
\begin{equation}\label{eq:phiL_calib}
\phi_\mathrm{L}=\phi_\mathrm{L/C}+\phi_\mathrm{C}(\mathrm{Model \,C1})= \phi_\mathrm{L/C},
\end{equation}
because Model~C1 is centre symmetric $\phi_\mathrm{C}(\mathrm{Model \,C1})=0$. This calibration has some limitations. Model~C1 overestimates the continuum squared visibility at small baselines (cf. Fig.~\ref{fig:continuum}). The line visibility at small baselines is slightly overestimated. With regards the phase, the stellar components could be moving during the orbit.

\paragraph*{Fit} The Br$\gamma$ visibility and phase are fitted with this model.  The data and model are presented in Fig.~\ref{fig:astrometry-jet-fit}. The best fit model parameters are presented in Table~\ref{tab:astrometry-jet-fit}. The knot has a very high flux ($F_\mathrm{knot}\sim3.5$~\AA), similar to values observed in a single emitting pre-main-sequence star. It is located very near the position of maximum astrometric displacement. It has no proper-motion. The star is positioned at an almost symmetrical distance from the centre of the field-of-view.

\paragraph*{Physical inconsistency of the fit} Although the model provides a reasonable fit to the data ($\chi^2_\nu=1.4$) it is physically inconsistent as it requires a too high $F_\mathrm{knot}$ and too small $v_\mathrm{knot}$.  Firstly, $F_\mathrm{knot}$ can be limited by an analysis of the line profile. As the system is seen almost pole-on, a significant flux in the knot would distort the line profile shape, with a blue-shifted enhanced contribution at the jet radial speed. Our analysis of the line profile limits  $F_\mathrm{knot}<1~$\AA\, (12-25~per cent of the total Br$\gamma$ flux), a much smaller value than the best-fitting $F_\mathrm{knot}$. The Br$\gamma$ line profile shape is similar to those commonly observed in  pre-main-sequence stars \citep{Folha2001,Garcia-Lopez2006,Najita1996}. Secondly, in the best fit model the knot is at rest, at odds with the proper-motion expected for a jet knot. \citet{Grady2004} measure for the Ly$\alpha$ knots HH\,669-A and HH\,669-B, a radial velocity of \mbox{$(-337\pm80)$~km~s$^{-1}$}. The corresponding speed in the plane of the sky would be \mbox{$v_\mathrm{\perp}=0.5   (v_\mathrm{jet}/\mathrm{360~km~s}^{-1})\,\mathrm{mas~d}^{-1}$}. A third difficulty is related to the fixed stellar position. The central object is a spectroscopic binary, both the longitude of the ascending node and the time of periastron passage would have to be fine-tuned to keep the star in the required position opposite to the jet direction.

We therefore rule out that a knot in the jet as the explanation for the compact Br$\gamma$ emission.

\begin{table}
\centering
\caption{Jet scenario best-fitting model parameters.}
\label{tab:astrometry-jet-fit}
\begin{tabular}{lc}
\hline
Parameter             & Value\\\hline
$F_\mathrm{knot}$ (\AA)      & $3.46\pm0.07$\\
$d_\mathrm{knot, t_0}$ (mas)& $1.0\pm0.2$\\
$v_\mathrm{knot}$ (mas~d$^{-1}$)  & $0.00\pm0.04$\\
$d_\star$ (mas)              & $1.0\pm0.6$\\
$\chi^2_\nu$                &1.4\\
\hline
\end{tabular}
\end{table}

\subsection{Binary scenario}
\label{sec:binary}

In this section we start by introducing the spectro-astrometric model of the data, then we present the $V_\mathrm{L/C}$ calibration to recover $V_\mathrm{L}$. The calibrated data is then fitted with orbital models in which the Br$\gamma$ flux is allowed to vary. The resulting variable emission is then discussed and interpreted as variable accretion.

\subsubsection{Spectro-astrometry}

Assume that only the primary and the secondary contribute to the total Br$\gamma$ emission and that the contribution is unresolved. The primary fractional emission is defined as
\[ \alpha_\mathrm{L} = \frac{ {\rm Br} \gamma_1}{ {\rm Br} \gamma_1 + {\rm Br} \gamma_2} \]
and the primary fractional mass as
\[ \alpha_m= \frac{m_1}{m_1 + m_2} .\]
It can be shown\footnote{The spectro-astrometric centre is $\bmath{c}=\alpha_\mathrm{L} \bmath{r}_1 + (1-\alpha_\mathrm{L}) \bmath{r}_2 $. The result is
obtained by substituting the two-body solutions (\mbox{$\bmath{r}_2=-\alpha_m \bmath{r}_{1/2}$} and \mbox{$\bmath{r}_1= (1-\alpha_m) \bmath{r}_{1/2}$}) in the
centre expression.} that the Br$\gamma$ spectro-astrometric centre ($\bmath{c}$) is given by
\[ \bmath{c}=(\alpha_\mathrm{L} - \alpha_m) \bmath{r}_{1/2},\]
where $\bmath{r}_{1/2}$ is the relative position of the primary with respect to the secondary.

The previous equation for the centre is degenerate.  The position angle of the orbit in the plane of the sky is determined by the longitude of the ascending node ($\Omega$). Orbits with opposing $\Omega$ can provide the same astrometric centre provided the stars fractional fluxes are adjusted. Quantitatively, the values
$\alpha_\mathrm{L}$ and $\Omega$ yield the same astrometric center\footnote{As long as $2\alpha_m-1<\alpha_\mathrm{L}<2\alpha_m$.} as the values $\alpha'_\mathrm{L}=2\alpha_m-\alpha_\mathrm{L}$ and $\Omega'=\Omega+\pi$. This degeneracy is broken by the use of the visibility ($V_\mathrm{L}$). Only in the very special case where $2\alpha_m=1$ is also the visibility symmetric. For HD\,104237 $2\alpha_m=1.2\pm0.1$ so both solutions are kept because they have similar $\chi^2_\nu$ (cf. Table~\ref{tab:binary-model}).

\subsubsection{Visibility calibration}\label{sec:alpha_c}

As for the jet scenario we have recovered calibrated line visibilities and phases making use of the continuum Model~C1 (cf. Eqs.~\ref{eq:VL_calib} and \ref{eq:phiL_calib}). This calibration does not address the fact that the continuum emission from the stars is moving during the orbital phase, creating a $\phi_\mathrm{C}\neq 0$. It can be shown that this effect is
\[ \bmath{c}_\mathrm{C}= (\alpha_1 - (\alpha_1+\alpha_2)\alpha_m)\bmath{r}_{1/2}=\alpha_\mathrm{C}\bmath{r}_{1/2},  \]
where $\alpha_i=F_{\mathrm{C},i}/F_\mathrm{T}$ is the $i^\mathrm{th}$ component continuum fractional flux and $F_\mathrm{T}=F_{\mathrm{C},1}+ F_{\mathrm{C},2}+ F_{\mathrm{C,disc}}$ is the total continuum flux. The net effect of the continuum centroid drift due to the orbital motion of the stars is to bias $\alpha_\mathrm{L}$with respect to the real one by roughly $\alpha_\mathrm{C}$. HD\,104237 has $\alpha_1=0.2$ and $\alpha_2=0.1$ \citep{Tatulli2007b}, thus $\alpha_\mathrm{C}\sim0.02$. A further fraction of $\sim 0.2$ of the total continuum is found unresolved in the continuum fits of Section~\ref{sec:LR-models}, in worst case it would change the previous value to $|\alpha_\mathrm{C}|\sim0.1$. This bias is generally smaller than our precision in $\alpha_\mathrm{L}$ and will only affect the fractional flux allocated to the binary components.

\subsubsection{Models}

\begin{figure*}
\begin{center}
  \includegraphics[angle=270, width=13cm]{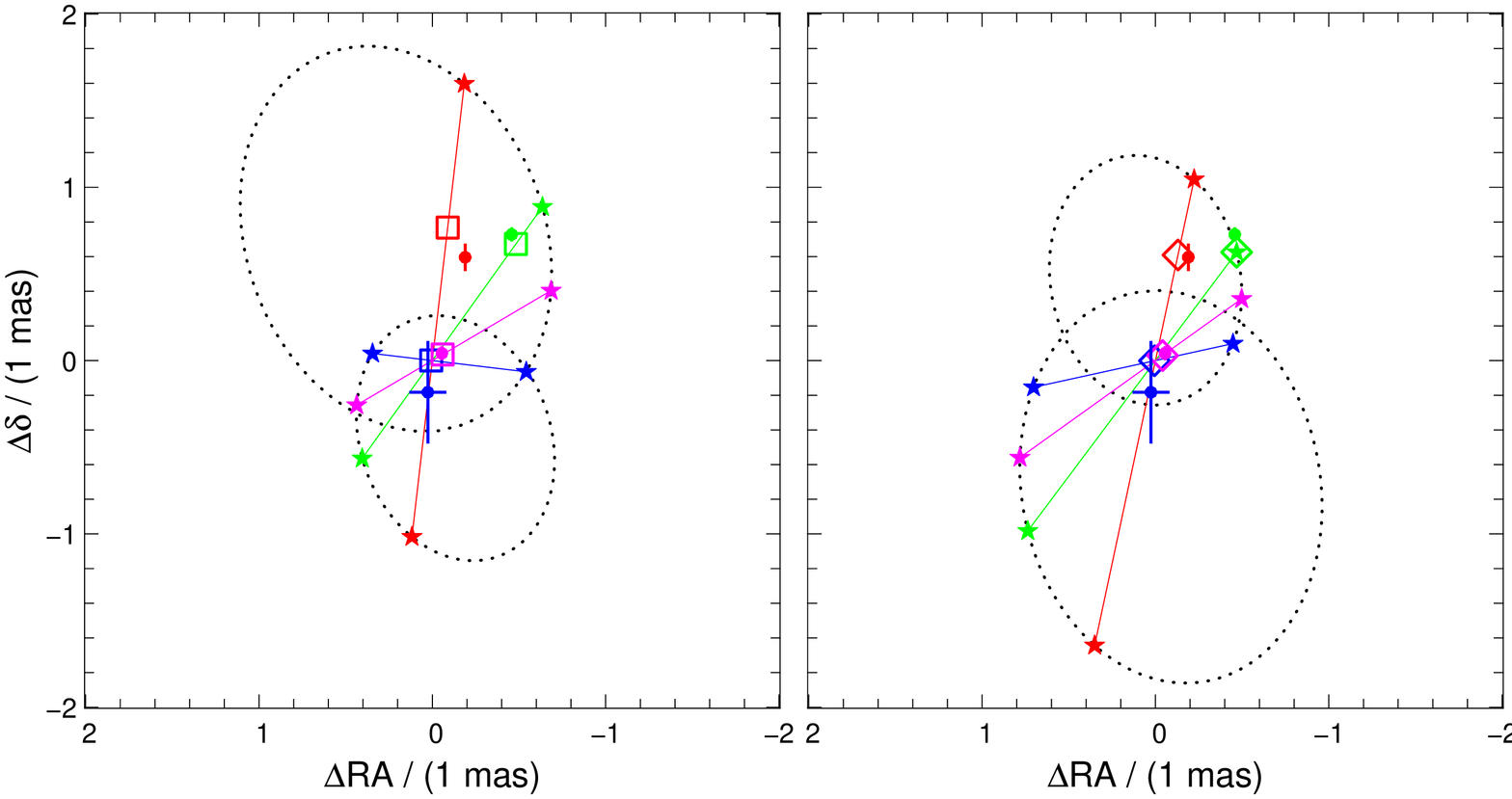}\\[10pt]
  \includegraphics[height=6.12cm]{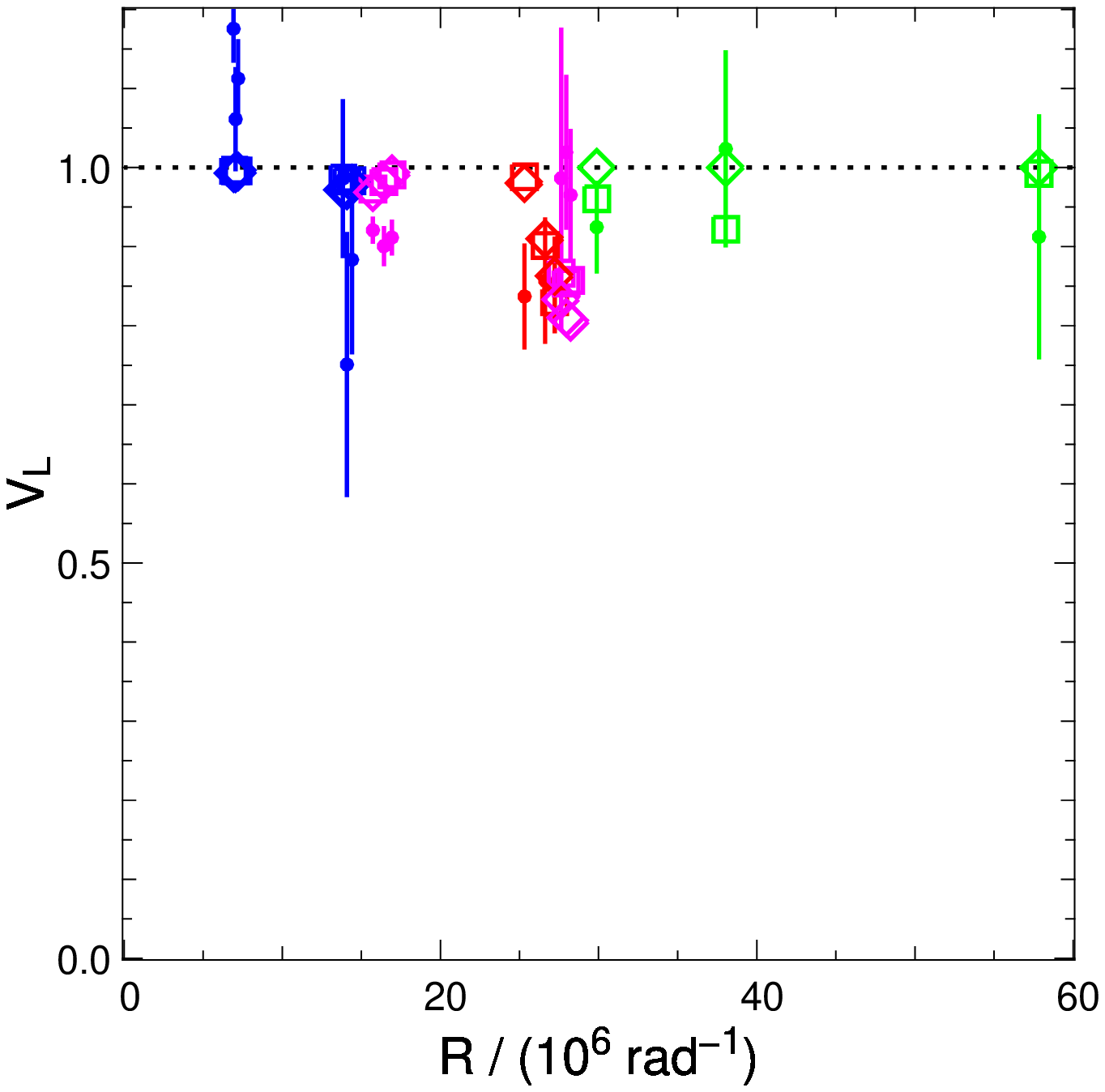}\hspace{5mm}
  \includegraphics[height=6.2cm]{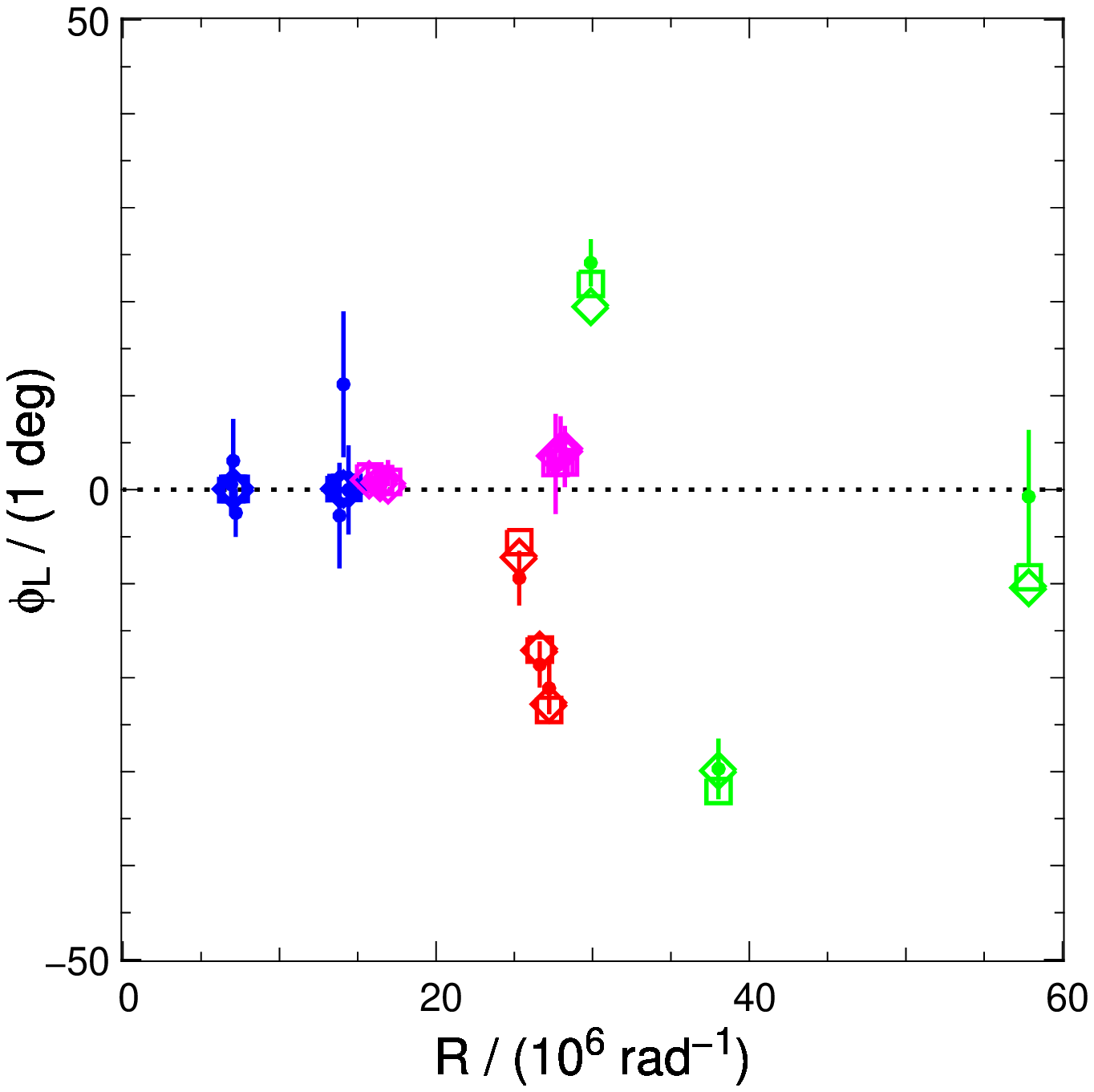}
\caption{\label{fig:binary-model} Retrograde binary scenario. Top: orbit drawing, solution with $(\Omega, \alpha_\mathrm{L})$ in the left, solution with $(\Omega',\alpha'_\mathrm{L})$ in the right. The dotted curves are the primary and secondary orbits. The stars ($\star$) are at the binary components positions at the time of observations. The crosses are the astrometric data. The open squares are the best-fitting model for the solution $(\Omega, \alpha_\mathrm{L})$, the open diamonds for the solution $(\Omega',\alpha'_\mathrm{L})$. Bottom: line visibility (left) and line phase (right) versus spatial frequency radius $R$ in the ($u,v$) plane.}
\end{center}
\end{figure*}

The model has six parameters: two orbital ($t_\mathrm{P,i}$, $\Omega$) and four for the nightly variable flux ratio $\alpha_\mathrm{L}$. The system orbit is computed using the  parameters in Table~\ref{tab:radial-velocity}. The model is fitted to the Br$\gamma$ visibilities and phases in  two steps. In the first, the full $\chi^2$ space is searched for a global minimum. In the second, the minimum is fine-tuned with a Levenberg-Marquardt algorithm.  Models with prograde and retrograde motion were fitted. The prograde models are a worse fit ($\chi^2_\nu=4.0$ and 4.8, for $\Omega$ and $\Omega'$, respectively) than the retrograde model (with $\chi^2_\nu=3.2$ and 3.4). The retrograde motion results are presented in Fig.~\ref{fig:binary-model} and Table~\ref{tab:binary-model}, they are our baseline model for the interpretation of the Br$\gamma$ emission from the system and are discussed in more detail in the next sections.

The average time of periastron passage obtained in the above retrograde fit ($t_\mathrm{P,i}$) is independent of the time of periastron passage obtained in the radial velocity fit ($t_\mathrm{P}$, Appendix~\ref{sec:stellar-parameters}). They are in very close agreement \mbox{$(t_\mathrm{P}-t_\mathrm{P,i})/P = 2601.004$}. The spectro-astrometric data can be used to determine a more accurate longitude of the ascending node by keeping $t_\mathrm{P,i}$ fixed at the radial velocity value in the above fits, obtaining $\Omega=(235\pm3)\degr$.

\begin{table*}
\centering
\caption{Retrograde binary scenario fit parameters.}
\label{tab:binary-model}
\begin{tabular}{lccccccc}
\hline

&\multicolumn{3}{c}{$\Omega=(234\pm8)\degr$} &\multicolumn{3}{c}{$\Omega'=(44\pm8)\degr$}&\\
&\multicolumn{3}{c}{$t_\mathrm{P,i}=(1.7\pm0.6)$~MJD} &\multicolumn{3}{c}{$t_\mathrm{P,i}'=(2.3\pm0.7)$~MJD}&\\
&\multicolumn{3}{c}{$\chi^2_\nu=3.2$} &\multicolumn{3}{c}{ $\chi^2_\nu=3.4$}&\\\hline

Night &$\phi_\mathrm{P}$  ($\degr$)&Br$\gamma$1 (\AA) & Br$\gamma$2 (\AA) &$\phi_\mathrm{P}$ ($\degr$) & Br$\gamma$1 (\AA) & Br$\gamma$2 (\AA) &Br$\gamma$T (\AA)  \\\hline
MR1 &255 &$-1.5 \pm 0.4$ &$-3.2\pm 0.9$ &244 &$-4.0\pm 0.7$ &$-0.8\pm 0.1$ &$-4.8$ \\
MR2 &308 &$-0.6 \pm 0.3$ &$-3.3\pm 2.0$ &298 &$-3.9\pm 0.6$ &$-0.0\pm 0.0$ &$-3.9$ \\
MR3 &346 &$-4.6 \pm 1.5$ &$-2.8\pm 0.9$ &335 &$-4.5\pm 1.3$ &$-2.9\pm 0.8$ &$-7.4$ \\
HR1 &330 &$-5.1 \pm 0.3$ &$-4.0\pm 0.2$ &319 &$-5.8\pm 0.3$ &$-3.2\pm 0.2$ &$-9.1$ \\\hline
\end{tabular}

\medskip
\raggedright

The table presents, for the two models labeled by $\Omega$ and  $\Omega'$,  the time of periastron passage ($t_\mathrm{P,i}$, $t_\mathrm{P,i}'$, in Modified Julian Date -- MJD), $\chi^2_\nu$, the position in the orbit ($\phi_\mathrm{P}$, periastron is $0\degr$) and each component (Br$\gamma$1,  Br$\gamma$2) and total (Br$\gamma$T) line equivalent width.

\end{table*}

\subsubsection{Br$\gamma$ variability}

The total Br$\gamma$ equivalent width (Br$\gamma$T) derived from the AMBER spectra is variable (cf. Table~\ref{tab:binary-model} and Fig.~\ref{fig:astrometry}). The interferometric data constrain the size (unresolved) and position of the Br$\gamma$ emission. The higher Br$\gamma$ emission (nights MR3 and HR1) takes place near periastron passage (cf. Table~\ref{tab:binary-model}). The fit extracts the individual components emission for each night which, as shown in Table~\ref{tab:binary-model}, is variable.  \citet{Donehew2011} refer to Br$\gamma$ line variability in their sample of Herbig Ae/Be stars of up to 4~\AA, when compared with the observations of \citet{Garcia-Lopez2006} and \citet{Brittain2007}. This equivalent width variability cannot be caused by the continuum. Roughly half of the continuum emission originates in the circumbinary disc inner rim whose Keplerian time-scales are $\sim70$~d,  much larger than the orbital period of the binary $\sim20$~d. From night MR2 to night MR3 (48\,h apart) a 3.5~\AA~(90\%) variation in the total Br$\gamma$ equivalent width is measured. Furthermore, \citet{Donehew2011} refer for their sample of Herbig Ae/Be stars a continuum variability of only $<10$~per cent. The pole-on geometry of HD\,104237 reduces shadowing effects and continuum variability.  The Br$\gamma$ equivalent width variability is most probably intrinsic to the line and originates in the binary stars very close environment.

\subsubsection{Accretion burst at periastron passage}

In classical T\,Tauri stars the Br$\gamma$ line could be formed in a magnetosphere, stellar wind or disc wind \citep[e.g.][]{Kurosawa2011, Kwan2011}. However, \citet{Kurosawa2011} don't find signatures of the disc wind in the Br$\gamma$ line, the bulk of the Br$\gamma$ emission takes place on scales smaller than 0.1~AU, mainly from the magnetosphere, and are unresolved by the present observations. This is agreement with \citet{Kraus2008} and \citet{Eisner2010} interferometric surveys of Br$\gamma$ who also find compact emission compatible with magnetospheric accretion in classical T\,Tauri stars and Herbig Ae stars.

Although the paradigm of magnetospheric accretion is less solid for Herbig Ae stars \citep[cf. discussion in][]{Donehew2011}, a correlation between accretion rate and Br$\gamma$ luminosity was found from classical T\,Tauri stars to Herbig Ae stars \citep{Muzerolle1998, Calvet2004, Donehew2011}. It can be argued that the Br$\gamma$ variability is at least an indirect tracer of variable accretion onto the stars. \citet{Garcia-Lopez2006} measured for HD\,104237 a Br$\gamma$T value of $\sim-5$~\AA~in 2004 and derived an accretion rate \mbox{$\dot\mathrm{M}_\mathrm{acc}\sim4\times10^{-8}~\mathrm{M}_\odot~\mathrm{yr}^{-1}$}. The exact fraction of the flux of each component could be slight biased by the continuum (cf. Section~\ref{sec:alpha_c}). For the fit with $\Omega=234\degr$, the primary average Br$\gamma$ flux more than doubles. For the fit with $\Omega'=44\degr$, it is the secondary Br$\gamma$ emission that has an even stronger increase. There is therefore at least a twofold increase in the accretion rate from nights MR1 and MR2 to the nights MR3 and HR1 (as indicated by the spectra alone). The later nights are those in which the system is nearer periastron. Indeed, if we take only the MR data (which is taken over the course of 5 nights and is more tightly correlated), the system is approaching periastron and an accretion burst from night MR2 to night MR3 would take place. The night HR1 high accretion rate measurement takes place one period after MR3, again near periastron (Fig.~\ref{fig:binary-model}). We cannot differentiate if there was a new burst of accretion at the second periastron passage or if the high accretion state remained during the full orbit.

\section{Accretion-ejection in HD\,104237}
\label{sec:accretion-ejection}

This section presents a synoptic view. The tidally disrupted circumbinary disc is initially addressed, followed by the accretion streamers and the interaction of the stellar magnetospheres. Finally, the origin of the Ly$\alpha$ jet is discussed.

\subsection{The tidally disrupted circumbinary disc}
\label{sec:inner_au}
\begin{figure}
\begin{center}
\includegraphics[width=0.8\columnwidth]{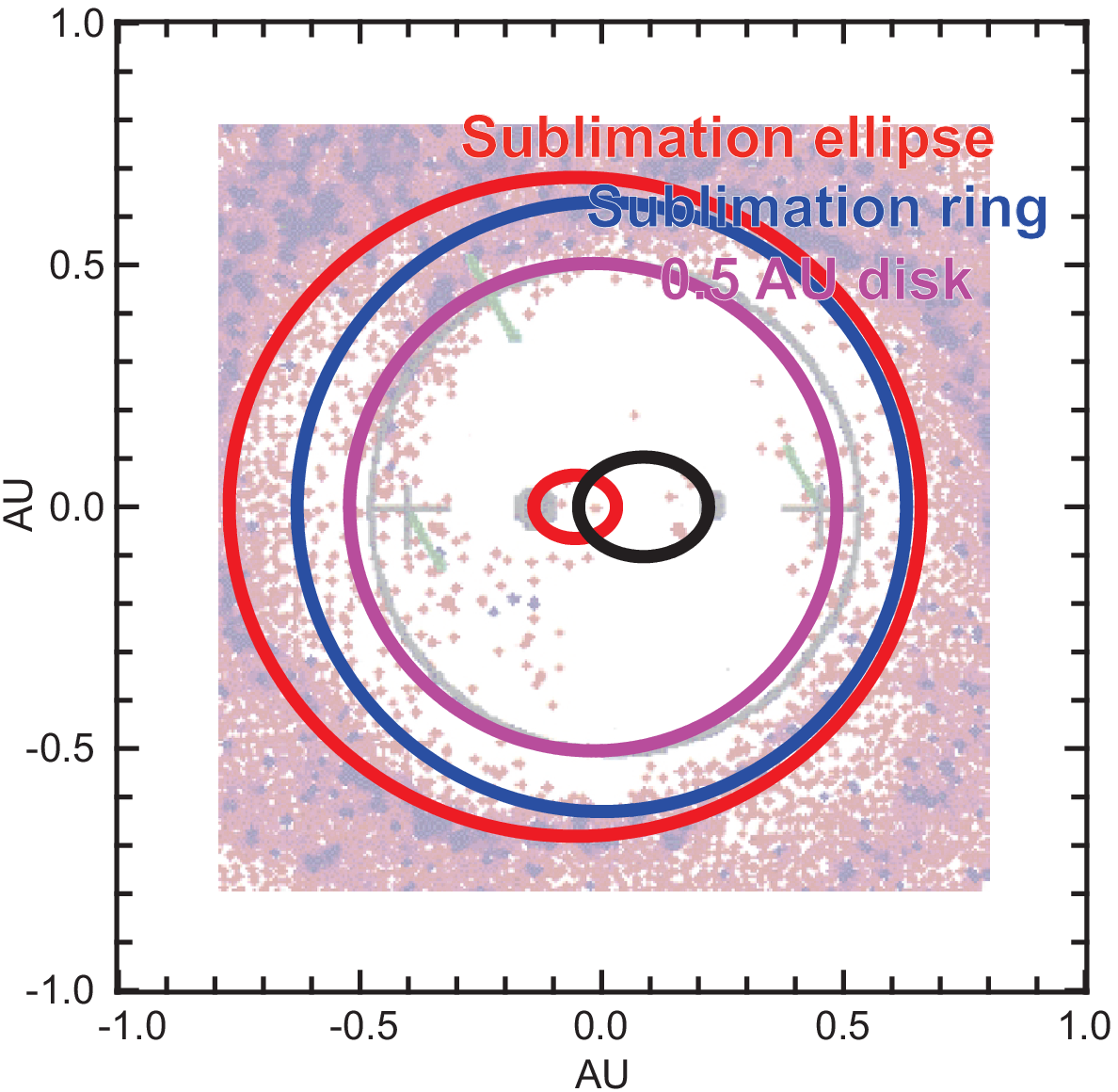}
\end{center}
\caption{\label{fig:binary_draw} Pole-on system diagram in apastron, with horizontal arbitrary position angle. The orbit of the primary is in red and of the secondary in black. The remaining curves trace, in increasing radius order:  a) 0.5~AU radius circumbinary disc (magenta); b) computed sublimation ring (blue); c) computed sublimation ellipse due to the primary orbit (red). The background is an image from \protect\citet{Artymowicz1996} at apocentre.  The image was scaled-up such that the stars are positioned over the orbits.}
\end{figure}

The tidally disrupted inner AU is schematically represented in Fig.~\ref{fig:binary_draw}. An inner gap is present and a circumbinary disc surrounds the system. The background image \citep[from][]{Artymowicz1996} presents a tidally truncated disc for a system with a mass ratio 1:1.27 and $e=0.5$. HD\,104237 has both a larger mass ratio (1:1.57) and eccentricity ($e=0.64$). Its truncation radius is expected to be located slightly further out than in the simulation, at   $\sim0.73$~AU using  \citet{Pichardo2008} scalings for HD\,104237 parameters. The squared visibility continuum models presented in Section~\ref{sec:LR-models} have radii smaller than the hydrodynamic truncation radius, the same applying to previous fits \citep{Tatulli2007b, Kraus2008}. They hint at additional material in the tidally truncated region, such as found in DQ\,Tau \citep{Boden2009, Carr2001}. This is further supported by the unresolved contribution in the continuum models being roughly 50~per cent, larger than the stellar 30~per cent derived in the SED fits by \citeauthor{Tatulli2007b}.

In steady state the dust is sublimated inside a ring of radius $r_\mathrm{sub}\sim 0.6$~AU \citep[e.g.][]{Dullemond2001, Isella2005}. It is depicted in Fig.~\ref{fig:binary_draw} and slightly smaller than the tidal truncation radius.  For the tidally disrupted disc of AK\,Sco, \citet{Alencar2003} also find that the sublimation radius is smaller than the truncation radius, i.e. the circumbinary disc rim has a temperature below the sublimation temperature. Due to the orbital movement of the primary, the effective sublimation radius could be larger and the dust sublimation "rim" elliptical (cf. Fig.~\ref{fig:binary_draw}), with an eccentricity of 0.36. The previous reasoning is only valid if the local cooling time ($t_{\rm cool})$ in the rim is larger than the orbital period ($P$). Following \citet{Nagel2010} the cooling time is estimated $t_{\rm cool}\sim 16$~d. As the primary orbits, the circumbinary disc rim region nearer the star will have a higher temperature than the regions further out.  This region will appear as a hot spot/arc that rotates around the disc inner rim, following the primary orbital motion. This hot spot/arc could also account for an excess of unresolved flux observed in the data.

\subsection{Accretion streamers}
\label{sec:acc_streamer}

\citet{Beck2012} detect enhanced H$_2$ emission  inside the tidally disrupted disc of GG\,Tau. The peak of H$_2$ emission is located at the inner edge of the circumbinary disc and coincident with a dust accretion streamer found by \citet{Pietu2011}. \citet{Folha2001} report drastic changes in the Br$\gamma$ emission of GG\,Tau. Streaming activity was also detected in younger systems \citep{Mayama2010}. High-resolution (magneto-)hydrodynamical models of accretion streams \citep[e.g.][]{Gunther2002, DeVal-Borro2011,Shi2012} find that the streams spiral from the circumbinary disc via the Lagrange points into the binary components. The mass accretion is variable but the streams are stable through the orbits. Our observations find that $\sim 20$~per cent of the total continuum is unresolved and not accounted by the stellar photospheres. The Br$\gamma$ data shows that $\gtrsim 90$~per cent of the emission is compact and therefore only a small fraction $\lesssim 10$~per cent can originate from a similar region where H$_2$ is observed in GG\,Tau or from large-scale streams. However, streaming material could contribute to the excess continuum and Br$\gamma$ emission in a shock further down the accretion stream, nearer the stars. The pole-on geometry would imply a very small radial velocity for the streaming gas and such a contribution (in contrast with the jet case) would pass unnoticed in the total Br$\gamma$ profile. \citet{Hanawa2010} find compression shocks in the region of the interaction of each circumstellar disc and transient hot spots in the primary circumstellar disc.  Such regions are also a good locus for compact Br$\gamma$ emission although, it is not obvious if they are directly applicable to HD\,104237 whose stars are expected to be devoid of discs as will be argued next.

\subsection{Interacting magnetospheres\label{sec:interacting}}

Tidal effects are expected to truncate the circumstellar discs from outside. Applying \citet{Pichardo2005} expressions to HD\,104237, upper limits of 4~R$_\odot$ for the primary disc and 3~R$_\odot$ for the secondary disc are obtained. On the other hand the stellar magnetic fields would truncate the circumstellar disks from the inside. The search for line-of-sight magnetic fields in HD\,104237 by \citet{Donati1997} and \citet{Wade2007} found marginal detections at a level of 50\,G. However, a 150~G dipolar field intensity\footnote{Taking \citeauthor{Garcia-Lopez2006} accretion rate and other relevant data in Appendix~\ref{sec:stellar-parameters}.} is required to truncate discs of such small radii \citep[e.g.][]{Bessolaz2008}. This difference could be reconciled by the system pole-on geometry, the dipolar magnetospheric loops would be reaching the photosphere almost perpendicular to the line-of-sight, the detected projected field being small with respect to the total dipolar field. Alternatively, there exists a strong field but it is multipolar or not well ordered and of difficult detection. \citet{Gregory2012} finds this is indeed the case for more massive pre-main-sequence stars with well developed radiative cores, such as HD\,104237. At the small radii of the putative circumstellar disks, multipolar fields can contribute significantly to the truncation and relax the required dipolar field to even lower values \citep{Adams2012}. Circumstellar discs are most probably untenable in each component.

At periastron the distance between the stars is 17~R$_\odot$.  \citet{Getman2008b} find that the fast rotating disc-less young stars have typical flares with sizes similar to the co-rotation radius. The size of the magnetospheres can be taken as the co-rotation radius. For the primary it is 12~R$_\odot$. The secondary could be a fast rotator (with a similar co-rotation radius as the primary) or a CTTS-like slow rotator with an even larger co-rotation radius. The magnetospheres of both stars are therefore expected to interact during periastron passage and flaring activity as observed in UZ\,Tau, DQ\,Tau, AK\,Sco or V4046\,Sgr would take place \citep[e.g.][]{Getman2011, Kospal2011, Salter2010}. These flares would liberate energy and increase the effective size of the system magnetosphere. If the Br$\gamma$ originates in this disordered magnetosphere, its flux would increase at periastron without important changes in line profile, as shown in Section~\ref{sec:binary} and observed in Fig.~\ref{fig:spectra}. This Br$\gamma$ line profile behaviour is similar to the one referred to by \citet{Basri1997} for the Balmer lines of DQ\,Tau.

\subsection{The Ly$\alpha$ jet}

\citet{Grady2004} detect a Ly$\alpha$ jet with a speed of $\sim340$~km~s$^{-1}$ in HD\,104237. In jet models where the launch takes place at the disc surface \citep[cf.][for a recent review]{Konigl2011} matter is strongly coupled to the magnetic field. For a given field line anchored at a cylindrical radius ($r_\mathrm{k}$) at the disc surface, most of the acceleration takes place up to the Alfv\'en radius ($r_\mathrm{A}$) and is measured by the ratio ($r_\mathrm{A}/r_\mathrm{k}$). If HD\,104237 Ly$\alpha$ jet is launched from the inner rim of the circumbinary disc, the ratio can be estimated $r_\mathrm{A}/r_\mathrm{k}\simeq 5$. This ratio is also related to the mass loss to mass accretion ratio and can be used to estimate the wind mass loss $\dot{M}_\mathrm{w}  \simeq 2\times 10^{-9}~\mathrm{M}_\odot~\mathrm{yr}^{-1}$. The magnetic field in the launching region can be estimated following \citet{Anderson2005} as $B_\mathrm{p}\sim 1$~G.  Such a high field cannot originate in a dipole scaling, but is typical of the fields required by models for the launching regions of jets in CTTSs. A centrifugal driven jet originating in the circumbinary disc is a plausible explanation of the Ly$\alpha$ jet.

A disc wind was introduced by \citet{Weigelt2011} to explain the Br$\gamma$ of the Herbig Be star MWC\,297.  But for this star the emission is resolved and not compact as we reported for HD\,104237 in Section~\ref{sec:models}. Such a model is not directly applicable to HD\,104237. The absence of significant extended Br$\gamma$ emission could be related to the tidal disruption of the circumstellar discs.

Models that rely on stellar magnetosphere-disc coupling, such as the X-wind \citep[e.g.][]{Cai2008} or disc-magnetosphere simulations \citep[e.g.][]{Lii2012}, are not directly applicable to HD\,104237 because spiraling material crossing the dynamically truncated gap is not expected to form stable circumstellar discs (as argued in Section~\ref{sec:interacting}). HD\,104237 poses a challenge to current disc-magnetosphere interaction models of jets. It would be of interest to extend the analysis of \citet{Shi2012} to jet formation.

The Ly$\alpha$ jet could be driven from the star. \textit{IUE} observations by \citet{Hu1991} find the \ion{Mg}{ii} h and k lines to have a strong P\,Cyg profile, with the absorption extending up to $\sim520$~km~s$^{-1}$. This is striking evidence of a powerful stellar wind from the polar regions of the primary. Interest in stellar winds has been revised after the suggestion that they tap on accretion energy \citep{Matt2005}. Detailed models for winds have been developed and applied to slow rotating CTTSs \citep{Cranmer2008, Cranmer2009} finding mass losses lower than observed. However, the HD\,104237 primary has a higher rotation rate and the expected mass loss would increase. On the other hand, the tidally perturbed accretion and the magnetospheres interaction at periastron could enhance the amount of energy dissipated in Alfv\'en waves and power the wind. An attractive scenario for HD\,104237 is one consisting of a stellar wind collimated by an outer circumbinary disc wind, as proposed by \citet{Sauty2011} for the intermediate mass star RY\,Tau.

\section{Conclusions}

Spectro-interferometric observations of HD\,104237 in the Br$\gamma$ line and adjacent continuum with $\lambda/2B=1.84$~mas maximum resolution are reported. It is found that the line is variable, presenting roughly a factor of two stronger equivalent width at periastron passage. At least 90\% of the the line emission is angularly unresolved. The spectro-astrometry of the line presents displacements with a position angle compatible with the jet. However, it is argued that emission does not originate in the jet but instead in the stars. It is found that modeling the spectro-astrometry and the angular size of the  Br$\gamma$ with a binary provides good agreement with observations. A mixed origin in a binary plus jet was not addressed due to the high number of parameters with regard to the available data. HD\,104237 corroborates that tight binaries of the Herbig Ae class present similar phenomena as well as similar challenges to the disc-magnetosphere interaction as their lower luminosity CTTSs counterparts.

\section*{Acknowledgments}

We thank the VLTI team at Paranal and J.-P.~Berger for taking part of the LR data. This research was partially supported by FCT-Portugal through Projects PTDC/CTE-AST/098034/2008, PTDC/CTE-AST/116561/2010 and by the European Commission Seventh Framework Programme under Grant Agreements 226604 and 237720. It has made use of the AMBER data reduction package of the Jean-Marie Mariotti Center (http://www.jmmc.fr/amberdrs). All the analysis was done with \textsc{yorick}, a free interactive data processing language written by David Munro (http://yorick.sourceforge.net/). The referee is thanked for suggestions which improved the paper.

\appendix

\section{Stellar parameters}
\label{sec:stellar-parameters}

\begin{table}
\begin{center}
\caption{\label{tab:radial-velocity} Fit of radial velocity data of \protect\citet{Bohm2004}.}
\begin{tabular}{lc}\hline
Physical quantity& Value\\\hline
$t_P$ ($\mathrm{HJD} - 2\,450\,000$~d)       & $1647.539 \pm 0.003$\\
$K_1$~(km~s$^{-1}$)       & $\phantom{0000}17.8 \pm 0.2\phantom{00}$      \\
$e$                & $\phantom{000}0.643 \pm 0.006$   \\
$\omega_{1/2}$ ($\degr$) & $\phantom{000}218.2 \pm 0.9\phantom{00}$\\
$P$~(d)         & $\phantom{00}19.856 \pm 0.002$  \\
$\gamma$~(km~s$^{-1}$)    & $\phantom{0000}14.1 \pm 0.1\phantom{00}$      \\
$\chi^2_\nu$          &\phantom{000}1.8\\\hline\\
\end{tabular}

\end{center}
\end{table}

The assumed parameters for the HD\,104237 system used throughout the paper are presented.  A fit to the \citet{Bohm2004} radial velocity data was done to derive error bars on the system orbital parameters, the results are presented in Table~\ref{tab:radial-velocity}. To derive the orbital semi-major axis the masses must be known.  The primary mass is taken as $2.2\pm0.2$~M$_\odot$. The mean value is the one determined by \citeauthor{Bohm2004} and \citet{Fumel2012}, by placing the primary in the HR diagram and using \citet{Palla2001} tracks. However, the error is increased here from 5 to 10~per cent, more in line with discussions of uncertainties of pre-main-sequence mass determinations \citep[e.g.][]{Blondel2006, Gennaro2012, Mathieu2007, Siess2001}. This is still smaller than the error assumed in the more recent analysis of \citeauthor{Fumel2012}, which probably overestimates it because of the use of mass-luminosity relations. The secondary mass was determined by \citeauthor{Bohm2004} to be 1.7~M$_\odot$, a value quite high for a K3 pre-main-sequence star \citep[e.g.][]{Mathieu2007}. Furthermore, such a high mass implies a system inclination angle of 15$\degr$. Using the Ly$\alpha$ jet detection information, \citeauthor{Grady2004} determine an inclination angle of $18^{+14\,\circ}_{-11}$. Using pseudo-periodic variability of the H$\alpha$ line and the projected rotational velocity of the system, \citet{Bohm2006} derive an inclination of $23^{+9\,\circ}_{-8}$, which implies a secondary mass of $\sim1.2$~M$_\odot$. On the other hand if we place the \citet{Bohm2004} secondary HR position in \citeauthor{Gennaro2012} tracks we obtain a  mass of $\lesssim1.4$~M$_\odot$, highlighting theoretical uncertainties. In this study, a secondary mass of $1.4\pm0.3$~M$_\odot$ is assumed. With the system mass fixed, the radial velocity data allows the determination of the remaining orbital parameters in Table~\ref{tab:physical-parameters}.

\begin{table}
\begin{center}
\caption{\label{tab:physical-parameters} Physical parameters of the system.}
\begin{tabular}{lcl}\hline
Physical quantity& Value & Reference\\\hline
$m_1$ (M$_\odot$)       & $\phantom{00}2.2 \pm 0.2\phantom{0}$  & \citealt{Bohm2004}\\
$m_2$ (M$_\odot$)       & $\phantom{00}1.4 \pm 0.3\phantom{0}$ & assumed\\
$a$ (AU)                & $\phantom{0}0.22 \pm 0.06$ & derived\\
$a$ (mas)               & $\phantom{00}1.9 \pm 0.6\phantom{0}$ & idem\\
$i$ ($\degr$)          & $17^{+12}_{-9}$ & idem \\
$T_\mathrm{eff,1}$ (K)    & $\phantom{.}8500\pm150\phantom{.}$ & \citealt{Fumel2012}\\
$L_{\star,1}$ (L$_\odot$)  & $35^{+5}_{-4}$ & \citealt{vandenAncker1998}\\
$R_{\star,1}$ (R$_\odot$)     & $\phantom{00}2.7\pm0.2$\phantom{0} & derived\\
$v_\mathrm{rot,1}\sin i$ (km~s$^{-1}$) & $\phantom{00.}12\pm2\phantom{00.}$ & \citealt{Donati1997}\\
$P_\mathrm{rot,1}$ (d)     & $\phantom{00}3.4\pm2.0\phantom{0}$ & derived\\
$F_\mathrm{C,1}/F_\mathrm{C,T}$ &0.2 & \citealt{Tatulli2007b}\\
$F_\mathrm{C,2}/F_\mathrm{C,T}$ &0.1 & ibid.\\\hline
\end{tabular}
\end{center}

\medskip

$F_\mathrm{C,1}/F_\mathrm{C,T}$ is the K-band primary stellar flux to total flux ratio.  $F_\mathrm{C,2}/F_\mathrm{C,T}$ is the secondary to total flux ratio, the remaining emission arises in the circumbinary disk.

\end{table}

For simplicity (and given the available data), full coplanarity is postulated, i.e. the stellar, orbital and disc inclination angle are assumed to be approximately the same. Spectro-astrometric observations of young binaries by \citet{Baines2006} and \citet{Wheelwright2011} support that the circumprimary disc is coplanar with the binary orbit. Observations of more evolved systems find orbital and stellar axis alignment for small separations \citep{Hale1994, Howe2009}. It is a plausible assumption for a system that probably formed via fragmentation of a common core.

\section{Least squares dispersed visibility deconvolution}\label{sec:least-squares}

The extraction of the line-to-continuum complex visibility from the
total to continuum complex visibility was addressed previously by
\citet{Weigelt2007}, but the effect of the spectrograph line spread
function was ignored. This is a good approximation when the line
is well resolved by the spectrograph and the SNR is high, which is not the case for our
observations. In this section we present a procedure to 'deconvolve'
the complex visibility from the effects of the spectrograph and
recover the line-to-continuum complex visibility.

\subsection{Spectral convolution in AMBER data}

The AMBER data consists of an interferometric channel and three
photometric channels, the four channels being spectrally dispersed
through a spectrograph. Carrying waves are fitted to the
interferometric channel, extracting the real $\mathfrak{R}(\lambda)$ and
imaginary $\mathfrak{I}(\lambda)$ parts of the coherent flux, at each
wavelength. Because the data is dispersed, the observables are
convolved with the spectrograph line spread function profile
$\mathcal{P}(\lambda)$. The interferometric observables are therefore
\[ \mathfrak{R}= \left(\{\sqrt{f_i f_j}  \bmath{v}_{ij}\} \ast \mathcal{P}\right)\cdot \bmath{r}\,\,\mathrm{and}\,\,
 \mathfrak{I}= \left(\{\sqrt{f_i f_j} \bmath{v}_{ij}\} \ast \mathcal{P}\right)\cdot \bmath{i},\]
where $\sqrt{f_i f_j} \bmath{v}_{ij}$ is the coherent flux vector (in the complex plane) for baseline $ij$ and $\bmath{r}$, $\bmath{i}$ are the versors (in the complex plane) along the real and imaginary axes. The detected photometric signal $F_i$ is also convolved by $\mathcal{P}$
\[ F_i= f_i \ast \mathcal{P}.\]
The above quantities are combined in the pipeline to yield the squared visibility\footnote{Details connected to debiasing the data are ignored as they are achromatic in the pipeline \citep{Tatulli2007a}.}
\[ V_{ij}^2 = \frac{\left|(\sqrt{f_i f_j} \bmath{v}_{ij}) \ast \mathcal{P} \right|^2}{F_i F_j}.\]
The above relation shows that the measured square visibility is not the convolution of the visibility by the spectrograph line spread function. The delivered differential phase observable is
\[ \Delta\phi_{\rm T/C}= \arctan\left(\frac{\mathfrak{I}}{\mathfrak{R}}\right)-\arctan\left(\frac{\mathfrak{I}_{\rm C}}{\mathfrak{R}_{\rm C}}\right), \]
which is also not a simple convolution of a real phase by the spectrograph line spread function. To invert the previous relations some model for the data has to be assumed.

\subsection{Spectral deconvolution method}

The spectra to continuum ratio in the photometric channel of telescope $i$ near Br$\gamma$ is described by
\[ F_{i/{\rm C}} =  1 + F_{\rm L/C} P_{\rm L},\]
where $ F_{\rm L/C}= F_{\rm L}/F_{\rm C}$ is the ratio of line amplitude to continuum value and $P_{\rm L}$ is the convolution of the intrinsic Br$\gamma$ profile $p_{\rm L}$ by the spectrograph line spread function.  The coherent flux vector is also assumed to be the combination of line and continuum contributions
\[ \sqrt{f_i f_j} \bmath{v}_{ij} = F_{\rm C} \bmath{V}_{\rm C} +  F_{\rm L} p_{\rm L} \bmath{V}_{\rm L}. \]
The line visibility vector (in the complex plane) $\bmath{V}_{\rm L}$ can change across the profile. However, after verification with the HR data, it is assumed to be constant for a given baseline.
With the above definitions it can be shown that
\begin{eqnarray}\label{eq:model:2}
V^2_{\rm T/C}&=&V_{ij}^2 / V_{\rm C}^2\\\nonumber
&=& \frac{1 + 2 F_{\rm L/C} P_{\rm L} V_{\rm L/C} \cos(\Delta\phi_{\rm L/C}) + F^2_{\rm L/C} P_{\rm L}^2 V^2_{\rm L/C}
}{F_{i/{\rm C}}  F_{j/{\rm C}} },
\end{eqnarray}
where $V_{\rm L/C}=V_{\rm L}/V_{\rm C}$, $V_{\rm T/C}=V_{\rm T}/V_{\rm C}$ and $\Delta\phi_{\rm L/C}=\phi_{\rm L}-\phi_{\rm C}$. The squared visibility normalized by the continuum $V^2_{\rm T/C}$ is independent of the interferometer transfer function and SNR biases and obtained in Section~2. However, we don't have direct access to the line-to-continuum visibility $V_{\rm L/C}$ and phase $\Delta\phi_{\rm L/C}$ \citep{Weigelt2007}. With regards to the differential phase
\begin{equation}\label{eq:model:3}
\tan(\Delta\phi_{\rm T/C})= \frac{V_{\rm L/C} F_{\rm L/C} P_{\rm L} \sin(\Delta\phi_{\rm L/C})}{1+V_{\rm L/C} F_{\rm L/C} P_{\rm L}\cos(\Delta\phi_{\rm L/C})}.
\end{equation}
In Eq.~\ref{eq:model:2} and Eq.~\ref{eq:model:3} the product $F_{\rm L/C} P_{\rm L}$ for a given baseline $ij$ can be obtained from the photometric channels data
\begin{equation}\label{eq:model:4}
F_{\rm L/C} P_{\rm L} = \sqrt{F_{i/{\rm C}} F_{j/{\rm C}}} -1.
\end{equation}
For a given baseline, Eqs.~\ref{eq:model:2} to \ref{eq:model:4} provide a complete description of the data with parameters $V_{\rm L/C}$ and $\Delta\phi_{\rm L/C}$, taking into account the spectrograph line dispersion function.

\subsection{Least squares fitting of the data}\label{sec:fit}

In practice, a Levenberg-Marquardt algorithm is used to fit the above model for the  $V^2_{\rm T/C}$ and $\tan(\Delta\phi_{\rm T/C})$ data, taking as \mbox{$x=F_{\rm L/C} P_{\rm L}$} given by Eq.~\ref{eq:model:4}. The  $\chi^2$ to be minimized is
\[\chi^2= \chi^2\{V^2_{\rm T/C}\}+\chi^2\{\tan(\Delta\phi_{\rm T/C})\}.\]

The errors in the weighting are constant and estimated from the
pseudo-continuum rms  in the $(2.13-2.15)~\mu$m range. Initial
estimates for the parameters are $V_{\rm L/C}=1$ and $\Delta\phi_{\rm
  L/C}=0\degr$, but the results are not sensitive to large variations of
initial estimates.  In the fit only points
within $|v| \leq 400$~km~s$^{-1}$ are used,  as including large amounts of
pseudo-continuum adjacent to the line would bias the results. Final
$\chi^2_\nu=1.3$ for the MR data and 1.4 for the HR data were obtained.

The fitted parameters are presented in Fig.~\ref{fig:fit-MR+HR}, the
actual visibility and phase models are presented in
Fig.~\ref{fig:data+fit+MR} and Fig.~\ref{fig:data+fit+HR}. Overall the
model is a very good description of the data, however there are
differential phases (MR1~G1-D0~06:42 and MR1~G1-D0~07:18, in
Fig.~\ref{fig:data+fit+MR}) which are slightly red-shifted with
respect to the line profile and squared visibility. This effect is
smaller than the error bars in the extracted quantities. The data set MR3~H0-G0~07:51 is apparently not well fit by the model,
it is the worst data in the set with a phase rms of $7^\circ$.

\subsection{Comparison with the differential method}

\begin{figure}
\begin{center}
\includegraphics[width=5cm]{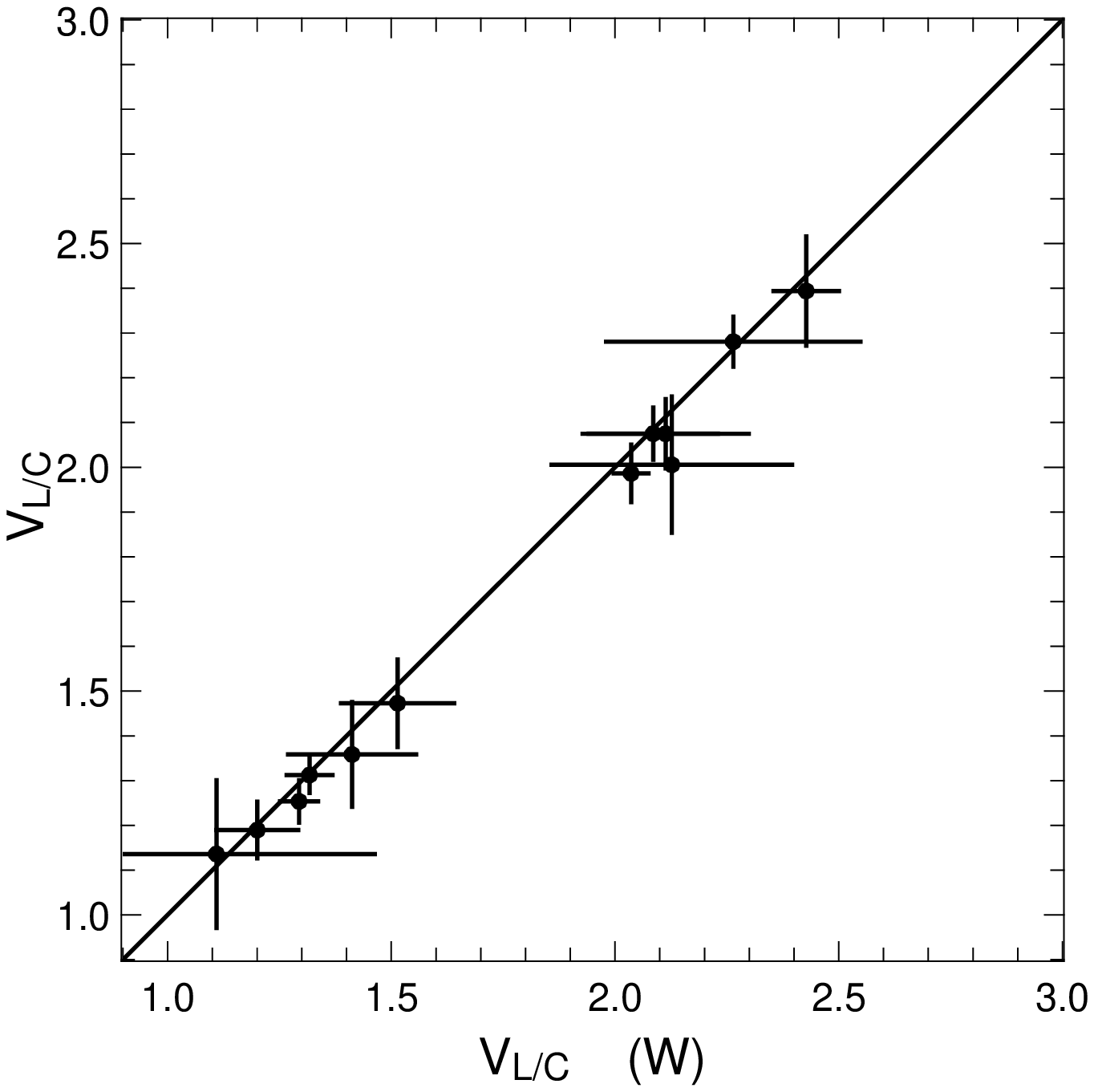}\\[10pt]
\includegraphics[width=5cm]{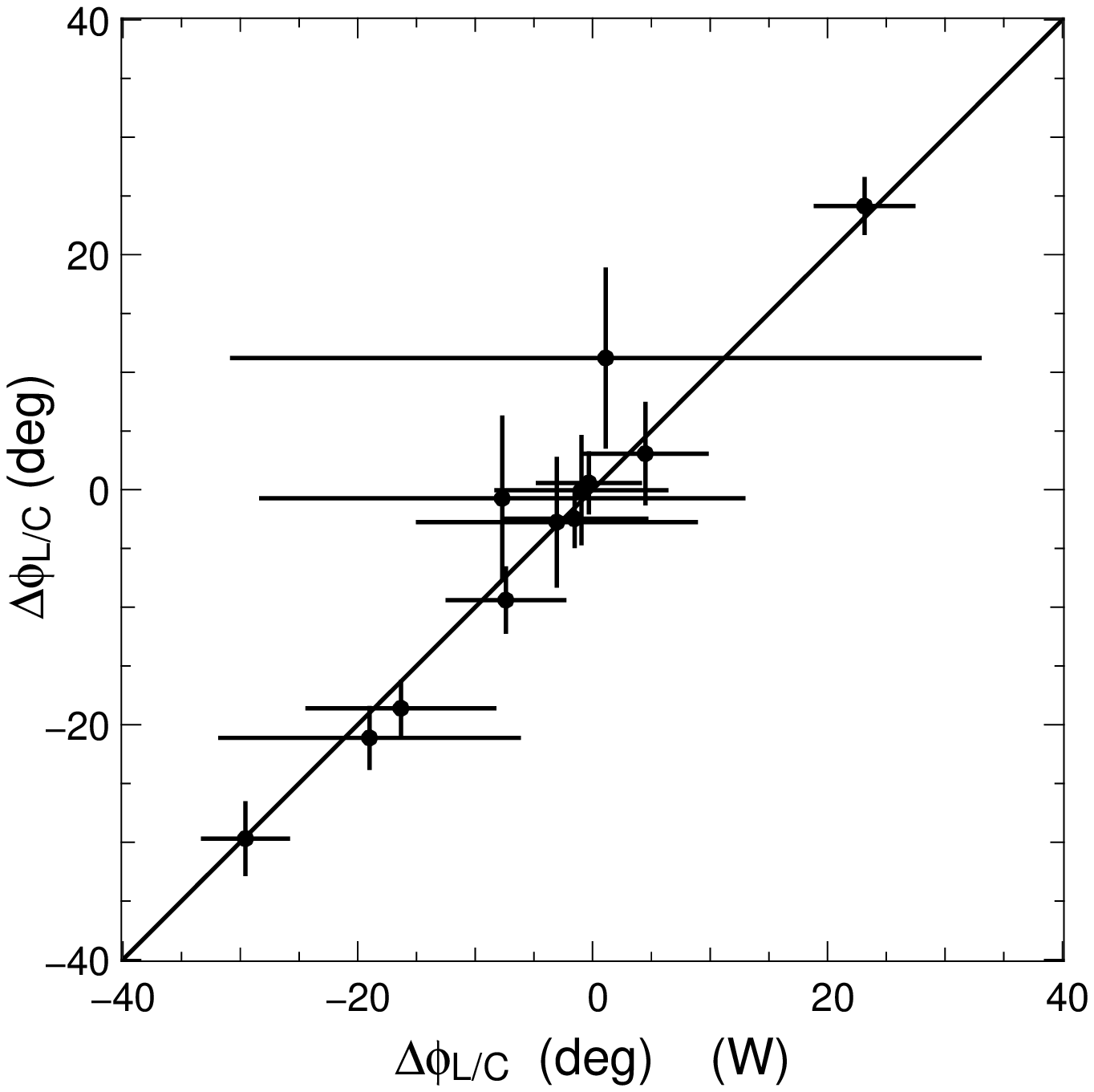}
\caption{\label{fig:comparison} Top: least squares  $V_{\rm L/C}$ versus differential value $V_{\rm L/C}$~(W). Bottom: least squares $\Delta\phi_{\rm L/C}$ versus differential value $\Delta\phi_{\rm L/C}$~(W).}
\end{center}
\end{figure}

The differential method of \citet{Weigelt2007} was also applied to our
data. The results of both methods are compared in
Fig.~\ref{fig:comparison}. The differential method presents a result
for each spectral pixel. In the comparison the average and standard
deviation of the best points, with $|v| \leq 150$~km~s$^{-1}$, are taken. It
is found that both methods present the same average value within
errors. However, the precision of the least squares deconvolution is
typically a factor of two better than that of the differential method.

\bibliographystyle{mn2e}
\bibliography{hd104237}


\bsp
\label{lastpage}
\end{document}